\documentclass[preprint,showpacs,preprintnumbers,amsmath,amssymb]{revtex4}

\usepackage{epsfig}
\usepackage{graphicx}
\usepackage{dcolumn}
\usepackage{bm}

\def\DESepsf(#1 width #2){\epsfxsize=#2 \epsfbox{#1}}

\begin{document}


\title{Charmless Three-body Baryonic $B$ Decays}

\author{Chun-Khiang Chua}
\author{Wei-Shu Hou}%
\author{Shang-Yuu Tsai}
\affiliation{%
Physics Department,
National Taiwan University,
Taipei, Taiwan 10764,
Republic of China
}%


\date{\today}

\begin{abstract}
Motivated by recent data on $B\to p\bar p K$ decay,
we study various charmless three-body baryonic $B$ decay modes, including
$\Lambda \bar p\pi$, $\Sigma^0\bar p \pi$, $p\bar p\pi$, 
$p\bar p \overline K {}^0$, in a factorization approach.
These modes have rates of order $10^{-6}$.
There are two mechanisms for the production of baryon pairs:
current-produced and transition.
The behavior of decay spectra from these baryon production mechanisms  
can be understood by using QCD counting rules.  
Predictions on rates and decay spectra can be checked in the near future.
\end{abstract}


\maketitle

\section{Introduction}

The Belle collaboration recently reported the observation of 
$B^-\to p \bar p K^-$ decay, the first ever charmless baryonic $B$ decay mode,
giving ${\mathcal B}=(4.3^{+1.1}_{-0.9}\pm0.5)\times10^{-6}$ \cite{Abe:2002ds}.
Three-body baryonic decay in $b\to c$ transitions have been 
observed \cite{Anderson:2000tz} previously, 
following a suggestion by Dunietz \cite{Dunietz:1998uz}.
It is interesting to compare the charmless case to the charmful one
and also to charmless two-body modes such as $B^0\to p\bar p$,
which has ${\mathcal B}<1.2\times10^{-6}$~\cite{Abe:2002er}.

It has been pointed out that reduced energy release 
(e.g. by a fast recoiling meson) would favor the generation of baryon pair 
and thus three-body baryonic modes could be 
enhanced over two-body rates \cite{Hou:2000bz}.
One of the signatures would be (baryon pair) threshold enhancement
in the three-body baryonic modes.
In our previous study of $B^0\to D^{*-}p\bar n$~\cite{Chua:2001vh},
we assumed factorization and obtained up to 60\% of experimental rate from
the vector current contribution.
The decay spectrum exhibits threshold enhancement.
The same threshold enhancement effect was predicted for the charmless 
$\rho p\bar n$ mode, giving ${\mathcal B}\sim 10^{-6}$ \cite{Chua:2001xn}.
It is interesting that the newly observed $p\bar p K$ mode shows 
such a threshold enhancement \cite{Abe:2002ds}.
With this encouragement we extend our study to charmless modes
such as $\Lambda\bar p\pi$, $\Sigma\bar p\pi$, $p\bar p\pi^-$,
$p\bar p K^-$ and $p\bar p \overline K {}^0$.
These modes are interesting not just by their (possibly) large rates,
but also by their accessibility. 
Some of these modes are studied in a recent work \cite{Cheng:2001tr}
that utilizes a factorization and pole model approach. 

In Sec. II, 
we extend the factorization approach to the charmless case, 
where one now has two mechanisms for baryon pair production.
In Sec. III, 
we discuss baryonic form factors and their associated
Quantum Chromo-Dynamics (QCD) counting rules \cite{Brodsky:1974vy}.
In Sec.~IV, 
the formulation is applied to the above mentioned charmless modes.
The threshold enhancement phenomenon is found to be 
closely related to the QCD counting rules.
Discussion and conclusion are given in Sec.~V,
while some useful formulas are collected in an appendix.

\begin{figure}[b!]
\epsfig{figure=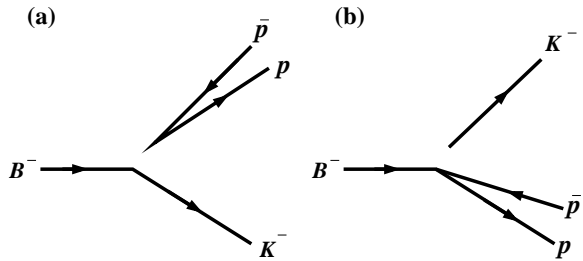,width=3in}
\caption{\label{fig:feynman} 
(a) The current-produced (${\mathcal J}$) and 
(b) transition (${\mathcal T}$) diagrams for $B^-\to p\bar p K^-$ decay.}
\end{figure}

\section{Factorization}\label{sec:fact}

In this section we extend the factorization approach used in 
Refs.~\cite{Chua:2001vh,Chua:2001xn} to charmless decay modes.
Under the factorization assumption, 
the three-body baryonic $B$ decay matrix element is separated into
either a {\it current-produced} baryon pair ($\mathcal J$) part
together with a $B$ to recoil meson transition part, or
a $B$ to baryon pair {\it transition} ($\mathcal T$)
part together with a current-produced recoil meson part.
As an example, the current-produced and 
transition diagrams for $B^-\to p\bar p K^-$ decay
are depicted in Fig. \ref{fig:feynman}.

\begin{table}[b!]
\caption{\label{tab:ai}The coefficients $a_i$ for $b\to
s~[b\to d\,]$ from \cite{Cheng:1999xj}. Values for
$a_3-a_{9}$ are in units of $10^{-4}$.}
\begin{tabular}{lccc}
\hline\hline
      & $N_c=2$ & $N_c=3$ & $N_c=\infty$ \\
\hline
$a_1$ & 0.99~[0.99] & 1.05~[1.05] & 1.17~[1.17] \\
$a_2$ & 0.22~[0.22] & 0.02~[0.02] & $-0.37~[-0.37]$ \\
\hline
$a_3$ & $-4.5 - 23\,i~[-2 - 20\,i]$ & $72.7 - 0.3\,i~[73 + 0.3\,i]$
                                  & $227 + 45\,i~[223 + 41\,i]$ \\
$a_4$ & $-349.5 -113.5\,i~[-338.5 -101.5\,i]$ & $-387.3 -121\,i~[-375.7 -108.3\,i]$
                                  & $-463 - 136\,i~[-450 - 122\,i]$ \\
$a_5$ & $-166 - 23\,i~[-164 - 20\,i]$ & $-66 - 0.3\,i~[-66 + 0.3\,i]$
                                  & $134 + 45\,i~[130 + 41\,i]$ \\
$a_6$ & $-533 - 113.5\,i~[-523 - 101.5\,i]$ & $-555.3 - 121\,i~[-544.7 - 108.3\,i]$ 
                                  & $-600 - 136\,i~[-588 - 122\,i]$ \\
$a_9$ & $-86.8 - 2.7\,i~[-86.6 - 2.5\,i]$ & $-92.6 - 2.7\,i~[-92.4 - 2.5\,i]$ 
                                  & $-104.3 - 2.7\,i~[-104.1 - 2.5\,i]$ \\
\hline\hline
\end{tabular}
\end{table}

In charmless decay modes, we need to use the effective Hamiltonian
consisting of operators and Wilson coefficients, which is standard and can
be found, for example, in Refs. \cite{Cheng:1999xj,Ali:1998eb}.
In this work, we concentrate on the dominant terms.
The factorization formula for the decay process $B\to X\,Y$, where 
$(X,Y)=(h,{\rm\bf B}\overline{\rm \bf B}^\prime)$ with $h$ being a light meson 
and ${\rm\bf B}\overline{\rm \bf B}^\prime$ a baryon pair or vice
versa is given by
\begin{eqnarray}
{\cal M}\left(B\to X\,Y\right) &=&\frac{G_F}{\sqrt2}\bigg\{ V_{ub}
V_{uq}^*[
   a_1(\bar q u)_{V-A}\otimes(\bar u b)_{V-A}
  +a_2(\bar u u)_{V-A}\otimes(\bar q b)_{V-A}]
\nonumber\\
&&-V_{tb} V_{tq}^*\Big[
   a_3\sum_{q^\prime}(\bar q^\prime q^\prime)_{V-A}\otimes(\bar q b)_{V-A}
  +a_4\sum_{q^\prime} (\bar q q^\prime)_{V-A}\otimes(\bar q^\prime
b)_{V-A}
\nonumber \\
&&\qquad
  +a_5\sum_{q^\prime}(\bar q^\prime q^\prime)_{V+A}\otimes(\bar q b)_{V-A}
 -2a_6\sum_{q^\prime} (\bar q q^\prime)_{S+P}\otimes(\bar q^\prime
b)_{S-P}
\nonumber\\
&&\qquad
  +\frac{3}{2} a_9\sum_{q^\prime} e_{q^\prime}
            (\bar q^\prime q^\prime)_{V-A}\otimes(\bar q b)_{V-A}
  +\dots
\Big]\bigg\}\,, 
\label{eq:factorization}
\end{eqnarray}
%
where ${\cal O}_1\otimes{\cal O}_2$ stands for $\langle X|{\cal
O}_1|0\rangle\langle Y|{\cal O}_2|B\rangle$. The coefficients
$a_i$ are defined in terms of the effective Wilson coefficients $c^{\rm eff}_i$ as
$a_{i={\rm odd}}\equiv c^{\rm eff}_i+c^{\rm eff}_{i+1}/N_c$ and 
$a_{i={\rm even}}\equiv c^{\rm eff}_i+c^{\rm eff}_{i-1}/N_c$.
We stress that $c^{\rm eff}_i$ are renormalization scale and scheme 
independent, as vertex and penguin corrections are included \cite{Cheng:1999xj}.
Their values are given in Table~\ref{tab:ai}.
In this work we use the $N_c=3$ case, while the $N_c=2,\infty$ cases are shown
to indicate non-factorizable effects.

\section{Form Factors and QCD Counting Rules}

In this section, we first discuss the meson form factors used in this work.
We then turn to discuss baryonic form factors, especially 
the implication of QCD counting rules.

\subsection{Meson Form Factors}

%
%

The decay constant $f_h$ of the pseudoscalar meson $h$ is defined
as
\begin{equation}
\langle h(p_h)|\bar q \gamma^\mu(1-\gamma_5)\,q^\prime|0\rangle
=i\, f_h \,p^\mu_h\,.
\end{equation}
These parameters and quark masses are taken from Ref. \cite {Ali:1998eb}.

We also need $0^-\to 0^-$ form factors defined as follows:
\begin{eqnarray}
\langle h\,(p_h)\left|\bar q \gamma^\mu(1-\gamma_5)\,b\right|
B(p_B)\rangle &=&
\biggl[(p_B+p_h)^\mu-\frac{m_B^2-m_h^2}{(p_B-p_h)^2}\,(p_B-p_h)^\mu
\biggr]\,F^{B\to h}_1(t)
\nonumber\\
&&\qquad\qquad\qquad
+\frac{m_B^2-m_h^2}{(p_B-p_h)^2}\,(p_B-p_h)^\mu\,F^{B\to h}_0(t)\,.
\end{eqnarray}
We use the so-called MS form factors, which take the
following form~\cite{Melikhov:2000yu}:
\begin{eqnarray}
F^{B\to h}_1(t)&=&\frac{F^{B\to h}_1(0)}
          {(1-t/M_V^2)[1-\sigma_1\,t/M_V^2+\sigma_2\,t^2/M_V^4]}\,,
\label{eq:f1}\\
F^{B\to h}_0(t)&=&\frac{F^{B\to h}_0(0)}
                       {1-\sigma_1\,t/M_V^2+\sigma_2\,t^2/M_V^4}\,,
\label{eq:f0}
\end{eqnarray}
where $M_V=5.42\ (5.32)$~GeV for $h= K(\pi)$. Other
parameters are given in Table~\ref{tab:meson_FF}.

\begin{table}[b!]
\caption{\label{tab:meson_FF}
Relevant parameters for the $B\to
K,\pi$ transition form factors of Eqs.~(\ref{eq:f1}) and
(\ref{eq:f0}).}
\begin{ruledtabular}
\begin{tabular}{lcccc}
  & $F^{B\to K}_1$ & $F^{B\to K}_0$ & $F^{B\to\pi}_1$ & $F^{B\to\pi}_0$ \\
\hline
$F_{1,0}^{B\to h}(0)$ & 0.36 & 0.36 & 0.29 & 0.29 \\
$\sigma_1$ & 0.43 & 0.70 &0.48 & 0.76 \\
$\sigma_2$ &  ---    & 0.27 &  ---   & 0.28 \\
\end{tabular}
\end{ruledtabular}
\end{table}

\subsection{Baryon Form Factors}

Factorization introduces two types of matrix elements containing the baryon pair: 
$\langle {\rm\bf B}\overline{\rm\bf B}^\prime|V(A)|0\rangle$ 
involving vector~($V$) or axial vector~($A$) {\it current-produced} baryon pair, 
and $\langle {\rm\bf B}\overline{\rm\bf B}^\prime|V(A)|B\rangle$
involving the $B\to {\rm\bf B}\overline{\rm\bf B}^\prime$ {\it transition}.

For the current-produced matrix elements, we have
\begin{eqnarray}
\langle {\rm\bf B}
\overline
{\rm \bf B}^\prime|V_{\mu}|0\rangle&=&
\bar{u}(p_{\rm\bf B}) \left\{F_1(t)\gamma_{\mu}+
   i\frac{\,F_2(t)}{m_{\rm\bf B}+m_{\overline{\rm\bf B}^\prime}}
\sigma_{\mu\nu}\left(p_{\rm\bf B}+p_{\overline{\rm\bf
B}^\prime}\right)^{\nu}\right\}v(p_{\overline{\rm\bf B}^\prime})
\nonumber\\
&=&\bar{u}(p_{\rm\bf B}) \left\{(F_1+F_2)\gamma_{\mu}+
   \frac{\,F_2(t)}{m_{\rm\bf B}+m_{\overline{\rm\bf B}^\prime}}
\left(p_{\overline{\rm\bf B}^\prime}-p_{\rm\bf B}\right)^\mu\right\}
v(p_{\overline{\rm\bf B}^\prime})\,,
\label{eq:vector_FF}\\
\langle {\rm\bf B}\overline{\rm\bf B}^\prime|A_{\mu}|0\rangle&=&
\bar{u}(p_{\rm\bf B})\left\{g_A\left(t\right)\gamma_{\mu}
+\frac{h_A\left(t\right)}{m_{\rm\bf B}+m_{\overline{\rm\bf
B}^\prime}} \,\left(p_{\rm\bf B}+p_{\overline{\rm\bf
B}^\prime}\right)_{\mu}\right\}\gamma_5\,
v(p_{\overline{\rm\bf B}^\prime})\,,
\label{eq:axial_FF}
\end{eqnarray}
where $F_{1,2}$ are the induced vector form factors, $g_A$
the axial form factor, and $h_A$ the induced pseudoscalar form factor. 
We have used Gordon decomposition to obtain 
the second line of Eq. (\ref{eq:vector_FF}). 
Note that $t\equiv (p_{\rm\bf B}+p_{\overline{\rm\bf B}^\prime})^2
  \equiv m^2_{{\mathbf B}\overline{\mathbf B}^\prime}$
is nothing but the ${\mathbf B}\overline{\mathbf B}^\prime$ pair mass.

According to QCD counting rules \cite{Brodsky:1974vy}, 
both the vector form factor $F_1$ and the axial form factor $g_A$,
supplemented with the leading logs, behave as $1/t^2$ in the $t\to\infty$ limit,
since we need two hard gluons to distribute large momentum transfer.
$F_2$ and $h_A$ behave as $1/t^3$, acquiring an extra $1/t$ due to helicity flip.
In the electromagnetic current case, the asymptotic form has been confirmed
by many experimental measurements of the nucleon magnetic (Sachs) form factor
$G_M=F_1+F_2$, over a wide range of momentum transfers in the space-like region.
The asymptotic behavior for $G^p_M$ also seems to hold in 
the time-like region, as reported by the Fermilab 
E760 experiment~\cite{Armstrong93} for $8.9$~GeV$^2<t<13$~GeV$^2$. 
Another Fermilab experiment, E835, 
has recently reported~\cite{Ambrogiani:1999bh} $G^p_M$ for 
momentum transfers up to $\sim 14.4$~GeV$^2$.
An empirical fit of
$
\left|G^p_M\right|=
C t^{-2} [\ln({t}/{Q_0^2})]^2
$,
is in agreement with the QCD counting rule prediction.

%
\begin{table}[t!]
\caption{\label{tab:formfactor}Relations of baryon form factors
$F_1+F_2$ and $g_A$ with the nucleon magnetic form factors
$G_M^{p,n}$.}
\begin{ruledtabular}
\begin{tabular}{lcccc}
${\mathbf B}\overline{\mathbf B}^\prime$
   &$V,\,A$ & SU(3) & $F_1+F_2$ &$g_A(t\to\infty)$\\
\hline $p\bar n$ &$(\bar u d)_{V,A}$ & $(F+D)_{V,A}$
          & $G^p_M-G^n_M$
          & 
            $\frac{5}{3}\, G^p_M+ G^n_M$
          \\
$\Lambda\bar p$ &$(\bar s u)_{V,A}$
          & $\left(-\sqrt{\frac{3}{2}}\,F-\sqrt{\frac{1}{6}}\,D\right)_{V,A}$
          & $-\sqrt{\frac{3}{2}}\,G^p_M$
          & 
            $-\sqrt{\frac{3}{2}}\,G^p_M$
           \\
$\Sigma^0\bar p$ & $(\bar su)_{V,A}$ &
          $ \frac{1}{\sqrt{2}}\,\left(D-F\right)_{V,A}$
          & $\frac{-1}{\sqrt{2}}\left(G_M^p+2\,G_M^n\right)$
          & 
            $\frac{1}{3\,\sqrt{2}}(G_M^p+6\,G_M^n)$
            \\
$p\bar p$ &$(\bar u u)_{V,A}$ & $(F+D+S)_{V,A}$
          &$G^p_M-G^n_M+S_V$
          & 
           $\frac{4}{3}\,G^p_M-G^n_M$
          \\
$p\bar p$ &$(\bar d d)_{V,A}$ & $S_{V,A}$
          &$S_V$
          & $-\frac{1}{3}\,G^p_M-2G^n_M$
          \\
$p\bar p$ &$(\bar s s)_{V,A}$ & $(D-F+S)_{V,A}$
          &$-G^p_M-2G^n_M+S_V$
          & 0
          \\
$p\bar p$ &$(\bar u u+\bar d d+\bar s s)_{V,A}$ & $(2\,D+3\,S)_{V,A}$
          &$3\,(-G^n_M+S_V)$
          & 
            $G^p_M-3\,G^n_M$
          \\
$p\bar p$ &$(e_u\,\bar u u+e_d\,\bar d d+e_S\,\bar s s)_{V,A}$ &
           $(F+D/3)_{V,A}$
          &$G^p_M$
          & 
            $G^p_M$
          \\
\end{tabular}
\end{ruledtabular}
\end{table}
%
%

The current induced form factors $F_1+F_2$ and $g_A$ can 
be related by means of the SU(3) decomposition form factors 
$F_{V,A}$, $D_{V,A}$ and $S_{V,A}$ (with $S_{V,A}$ appearing only in the 
non-traceless current case), as shown in Table~\ref{tab:formfactor}.
It is well known that $F_V$ and $D_V$ can be expressed by the 
nucleon magnetic form factors $G_M^{p,n}$,
\begin{eqnarray}
F_V = G_M^p+\frac{1}{2}\,G_M^n\,, \ \ \ \ \
D_V = -\frac{3}{2}\,G_M^n.
\label{eq:FVDV}
\end{eqnarray}
As the first term $F_1+F_2$ in Eq. (\ref{eq:vector_FF}) can be related to
nucleon magnetic Sachs form factor $G_M$,
similarly the second term $F_2$ can be related to
$(G_E-G_M)/[t/(m_{\bf B}+m_{\bf B^\prime})^2-1]$, 
where $G_E$ is the nucleon electric Sachs form factor.
Since we do not have enough data on time-like nucleon $G_E$,
we concentrate on the $F_1+F_2$ term as we did in Ref. \cite{Chua:2001vh}.
We may in fact gain information on $G_E$ by reversing our present analysis
on these three-body baryonic $B$ decays in the future when more data 
become available.

The nucleon magnetic form factors are fitted to available data in
Ref.~\cite{Chua:2001vh} by
\begin{eqnarray}
G_M^p(t)=\sum^5_{i=1}\frac{x_i}{t^{i+1}}
\left[\ln\left(\frac{t}{\Lambda_0^2}\right)\right]^{-\gamma}&,&
G_M^n(t)=-\sum^2_{i=1}\frac{y_i}{t^{i+1}}
\left[\ln\left(\frac{t}{\Lambda_0^2}\right)\right]^{-\gamma}\,,
\label{eq:GM}
\end{eqnarray}
where $\gamma=2.148$,
$x_1=420.96$~GeV$^4$, $x_2=-10485.50$~GeV$^6$,
$x_3=106390.97$~GeV$^8$, $x_4=-433916.61$~GeV$^{10}$,
$x_5=613780.15$~GeV$^{12}$, $y_1=236.69$~GeV$^4$,
$y_2=-579.51$~GeV$^6$, and $\Lambda_0=0.3$~GeV.
They satisfy QCD counting rules and describe time-like electromagnetic data 
such as $e^+e^-\to N \overline N$ suitably well.
We have real and positive (negative) time-like $G^{p(n)}_M$ 
\cite{Hammer:1996kx,Baldini:1999qn}.
It is interesting to note the alternating signs of
the $x_i$ and $y_i$ parameters,
and that only two terms are needed to describe the neutron 
magnetic form factor \cite{Chua:2001vh}.

The time-like form factors related to $S_V$, $F_A$, $D_A$, $S_A$ 
are not yet measured. 
It is noted in Ref. \cite{Cheng:2001tr} that the asymptotic behavior of
baryon form factors studied in the 80s may be useful.
Their asymptotic behavior
as $t\to\infty$ can be described by two form factors
depending on the reacting quark having parallel or anti-parallel
spin with respect to baryon spin~\cite{Brodsky:1980sx}.
By expressing these two form factors in terms of $G_M^{p,n}$ as
$t\to\infty$, one has
\begin{eqnarray}
S_V &\to& G_M^p+2\,G_M^n,
\nonumber\\
F_A &\to& \frac{2}{3}\,G_M^p-\frac{1}{2}\,G_M^n,
\nonumber\\
D_A &\to& G_M^p+\frac{3}{2}\,G_M^n,
\nonumber\\
S_A &\to& -\frac{1}{3}\,G_M^p-2\,G_M^n.
\label{eq:asymptotic}
\end{eqnarray}
Since these relations only hold for large $t$,
it implies relations on the leading terms of these form factors.
In general more terms may be needed.
In analogy to the neutron magnetic form case, 
we express these form factors up to the second term
\begin{eqnarray}
S_V(t)& \equiv & \left(\frac{s_1}{t^2}+\frac{
s_2}{t^3}\right)
\left[\ln\left(\frac{t}{\Lambda_0^2}\right)\right]^{-\gamma},
\nonumber\\
F_A(t) & \equiv & \left(\frac{\tilde f_1}{t^2}+\frac{\tilde
 f_2}{t^3}\right)
 \left[\ln\left(\frac{t}{\Lambda_0^2}\right)\right]^{-\gamma},
\nonumber\\
D_A(t)& \equiv & \left(\frac{\tilde d_1}{t^2}+\frac{\tilde
 d_2}{t^3}\right)
 \left[\ln\left(\frac{t}{\Lambda_0^2}\right)\right]^{-\gamma},
\nonumber\\
S_A(t)& \equiv & \left(\frac{\tilde s_1}{t^2}+\frac{\tilde
s_2}{t^3}\right)
\left[\ln\left(\frac{t}{\Lambda_0^2}\right)\right]^{-\gamma}.
\label{eq:FF}
\end{eqnarray}
The asymptotic relations of Eq. (\ref{eq:asymptotic}) imply
$s_1=x_1-2y_1$,
$\tilde f_1 = 2\,x_1/3+y_1/2$, 
$\tilde d_1 = x_1-3\,y_1/2$
and $\tilde s_1 = -x_1/3+2\,y_1$.

The coefficients of the second terms are undetermined due to
the lack of data. 
However, we can use the axial vector ($g^{pn}_A=F_A+D_A$) 
contribution in $B^0\rightarrow D^{*-}p\bar{n}$ decay
to constrain $\tilde f_2$ and $\tilde d_2$.
The part of the branching fraction
${\cal B}(B^0\rightarrow D^{*-}p\bar{n}) =
(14.5^{+3.4}_{-3.0}\pm2.7)\times 10^{-4}$~\cite{Anderson:2000tz}
arising from the vector current has been calculated to
give ${\mathcal B}_V\sim 7\times 10^{-4}$~\cite{Chua:2001vh}. 
We find for $\tilde f_2+\tilde d_2=-2110\ {\rm GeV}^6$, 
the branching fraction coming from the axial current 
${\cal B}_A(B^0\rightarrow D^{*-}p\bar{n})\sim 12.7\times 10^{-4}$, 
and the sum ${\cal B}_V+{\cal B}_A$ is within the measurement range. 
Had we used the asymptotic form of Eq.~(\ref{eq:asymptotic}) 
for $g^{pn}_A$ in the whole time-like region, 
we would obtain ${\cal B}_A\sim1.0\times10^{-4}$, which is too small.

In this work we take 
\begin{equation}
\tilde f_2+\tilde d_2\equiv \tilde z=-2110~{\rm
GeV}^6\,, \label{eq:constraint}
\end{equation}
and $s_2=\tilde s_2=0$ for simplicity, 
since there is no data to constrain these yet.
It is interesting to note that the asymptotic relations give vanishing
results for 
$\langle p\bar p| (\bar s s)_{V,A}|0\rangle\ (\sim D_{V,A}-F_{V,A}+S_{V,A})$,
as one can see from Table~\ref{tab:formfactor} and Eq.~(\ref{eq:asymptotic}).
This can be understood as OZI suppression.
We still have a vanishing $\langle p\bar p| (\bar s s)_{A}|0\rangle$ 
for smaller~$t$, 
if we take $\tilde d_2=\tilde f_2=\tilde z/2$.
Since there is no point to advocate a large $\bar s s$
form factor in this work and this choice is preferred by the OZI rule, 
we therefore use $\tilde d_2=\tilde f_2=\tilde z/2$ throughout.
For the vector case, we have vanishing 
$\langle p\bar p| (\bar s s)_{V}|0\rangle$ if we use the asymptotic relation
in the whole time-like region.
On the other hand, if we use Eq. (\ref{eq:FF}) for $S_V$, we may have 
a small but non-vanishing $(\bar s s)_V$ form factor for small $t$.
This may be related to the $\phi$ pole effect in the VMD view point 
\cite{Meissner:1996fn}.
Furthermore, we find that other OZI suppressed current-produced matrix elements,
such as 
$\langle n\bar n|(\bar s s)_{V,A}|0\rangle$,  
$\langle \Sigma^+\overline \Sigma {}^-|(\bar d d)_{V,A}|0\rangle$,
$\langle \Sigma^-\overline \Sigma {}^+|(\bar u u)_{V,A}|0\rangle$,
$\langle \Xi^-\overline\Xi {}^+|(\bar u u)_{V,A}|0\rangle$ and 
$\langle \Xi^0\overline\Xi {}^0|(\bar d d)_{V,A}|0\rangle$,
have the same SU(3) decomposition as 
the $\langle p\bar p| (\bar s s)_{V,A}|0\rangle$ one.
They therefore do not provide any further constraint.

We will also encounter 
$\langle{\rm\bf B}\overline{\rm\bf B}^\prime|\bar q q^\prime|0\rangle$,
which can be related to the vector matrix element by the equation of motion,
\begin{eqnarray}
\langle{\rm\bf B}\overline{\rm\bf B}^\prime|\bar
q q^\prime|0\rangle
&=&
\frac{\left(p_{\rm\bf B}+p_{\overline{\rm\bf B}^\prime}
\right)^\mu}{m_q-m_{\bar q^\prime}}
\langle{\rm\bf B}\overline{\rm\bf B}^\prime|V_\mu|0\rangle
\nonumber\\
&=& \frac{m_{\rm\bf B}-m_{\overline{\rm\bf B}^\prime}}{m_q-m_{\bar q^\prime}}\,
F_1(t)\bar u(p_{\rm\bf B}) v(p_{\overline{\rm\bf B}^\prime}).
\label{eq:0toBB}
\end{eqnarray}
This gives safe chiral limit in the ${\rm\bf B}\neq{\rm\bf B}^\prime$ case.
For example, in $\langle \Lambda\bar p|\bar u s|0\rangle$, we have
$({m_{\Lambda}-m_{\bar{p}}})/({m_s-m_{\bar u}})\sim 1$.
If ${\rm\bf B}={\rm\bf B}^\prime$, 
we encounter $({m_{\rm\bf B}-m_{\overline{\rm\bf B}}})/({m_q-m_{\bar q}})$.
As hinted from the $\Lambda \bar p$ case,
our ansatz is to take this factor as the number $n_q$ of the corresponding
constituent quark in $\bf B$.
For example, we take  
$\langle\bar p p|\bar d d|0\rangle\sim F_1\bar u v$, while
$\langle\bar p p|\bar s s|0\rangle\sim 0$ as suggested by the OZI rule.

For $h_A$, we follow Ref. \cite{Cheng:2001tr} to control the behavior
of pseudoscalar form factors in the chiral limit by using
\begin{equation}
h_A(t)=-\frac{(m_{\rm\bf B}+m_{\overline{\rm\bf B}^\prime})^2}
       {t-m^2_{\rm GB}}\, g_A(t)\,,
\label{eq:PCAC}
\end{equation}
where $m_{\rm GB}$ stands for the corresponding Goldstone boson mass.
Thus, in the chiral limit,
\begin{eqnarray}
\langle{\rm\bf B}\overline{\rm\bf B}^\prime|\bar
q\gamma_5
q^\prime|0\rangle
&=&
\frac{\left(p_{\rm\bf B}+p_{\overline{\rm\bf B}^\prime}
\right)^\mu}{m_q+m_{q^\prime}}
\langle{\rm\bf B}\overline{\rm\bf B}^\prime|A_\mu|0\rangle
\nonumber\\
&=& -\frac{m_{\rm GB}^2}{m_q+m_{q^\prime}}
\frac{m_{\rm\bf B}+m_{\overline{\rm\bf B}^\prime}}{t-m^2_{\rm GB}}\,
g_A(t)\bar u(p_{\rm\bf B})\gamma_5
v(p_{\overline{\rm\bf B}^\prime})
\label{eq:0toBBPCAC}
\end{eqnarray}
stays finite, otherwise, 
we will be facing a large enhancement factor
$({m_{\rm\bf B}+m_{\overline{\rm\bf B}^\prime}})/({m_q+m_{q^\prime}})$
in the above equation as we turn off $h_A(t)$.

We now turn to the transition form factors.
In general, the matrix element of 
$B\to {\rm\bf B}\overline{\rm\bf B}^\prime$ transition 
can be defined as
\begin{equation}
\langle{\rm\bf B}\overline{\rm\bf B}^\prime |\bar q
\left(\gamma_5\right)b
|B \rangle \equiv i\,\bar u(p_{{\rm\bf B}})
\left[{\cal F}_{V(5)}\,\,/\!\!\!p_h+{\cal F}_{A(5)}\,\,/\!\!\!p_h\gamma_5
     +{\cal F}_{P(5)}\,\gamma_5+{\cal F}_{S(5)}
\right] v(p_{\overline{\rm\bf B}^\prime}),
\label{eq:BBB_FF}
\end{equation}
where $p_h\equiv p_B-p_{\rm\bf B}-p_{{\rm\bf B}^\prime}$, 
and the form factors
${\cal F}_{V}={\cal F}_{A5}={\cal F}_{S}={\cal F}_{P5}=0$ 
by parity invariance.
According to QCD counting rules, 
for $t=m^2_{{\mathbf B}\overline{\mathbf B}^\prime}\to\infty$, we
need {\it three} hard gluons to distribute the large momentum
transfer released from the $b\to q$ transition. 
An additional gluon kicks the spectator quark in the $B$ meson
such that it becomes energetic in the final baryon pair. Thus, as
$t\to\infty$, we have:
\begin{equation}
{\cal F}_{A,V5}\to \frac{1}{ t^{\,3}}\,,\qquad 
{\cal F}_{P,S5}\to \frac{1}{t^{\,4}}\,.\label{eq:F_asymp}
\end{equation}
That ${\cal F}_{P,S5}$ have one more power of $1/t$ 
than ${\cal F}_{A,V5}$ is due to helicity flip.
This can be easily seen by taking 
$|\overline B_q\rangle\sim \bar b \gamma_5 q|0\rangle$. 
Without any chirality flip due to quark masses, 
we only have ${\cal F}_{A,V5}$ and the above counting rule holds,
while with additional chirality flip, more effectively from the $b$ quark mass,
we can also have ${\cal F}_{P,S5}$ but with additional power of $1/t$.

In this work we will need the transition matrix elements
$\langle p \bar p|(\bar u b)_{S,P}|B^-\rangle$ and 
$\langle p \bar p|(\bar d b)_{S,P}|\bar B^0\rangle$, 
which consist of eight form factors in total.
It is useful to restrict these even if only by some asymptotic relations.
By following a similar path to 
Ref. \cite{Brodsky:1980sx},
the chiral conserving parts ${\cal F}_{A,V5}$ can be expressed by 
two form factors depending on the interacting quark having parallel 
or anti-parallel spin with the proton spin.
The chiral flipping parts ${\cal F}_{P,S5}$ can be expressed by one
form factor with the spin of interacting quark parallel to the proton's.
The spin anti-parallel part is absent since it corresponds to an octet-decuplet 
instead of an octet-octet baryon pair final state.
The asymptotic forms (as $m^2_B,\,t\to \infty$) are
\begin{eqnarray}
\langle p \bar p |(\bar u b)_S|B^- \rangle 
&=&i\,\bar u(p_p)
  \left[F_{A}\,\,/\!\!\!p_h\gamma_5+F_{P}\,
  \gamma_5\right] v(p_{\overline p}),
\nonumber\\
\langle p \bar p |(\bar u b)_P|B^- \rangle 
&=&i\,\bar u(p_p)
  \left[F_{V5}\,\,/\!\!\!p_h+F_P\right] v(p_{\overline p}),
\nonumber\\
\langle p \bar p |(\bar d b)_S|\overline B {}^0 \rangle 
&=&i\,\bar u(p_p)
  \left[\frac{1}{10}(11 F_A+9 F_{V5})\,\,/\!\!\!p_h\gamma_5-\frac{1}{4}\,F_P\,
  \gamma_5 \right] v(p_{\overline p}),
\nonumber\\
\langle p \bar p |(\bar d b)_P|\overline B {}^0 \rangle 
&=&i\,\bar u(p_p)
  \left[\frac{1}{10}(9 F_A+11 F_{V5})\,\,/\!\!\!p_h-\frac{1}{4}\,F_{P}\right] 
v(p_{\overline p}).
\label{eq:ppB_FF}
\end{eqnarray}  
We need only three form factors.
For {\it simplicity}, we use
\begin{equation}
F_{A,V5}=\frac{C_{A,V5}}{t^{\,3}}\,,\qquad
F_{P}=\frac{C_P}{t^{\,4}}\,,
\end{equation}
and Eq. (\ref{eq:ppB_FF}) in the whole time-like region.

\section{Charmless Baryonic $B$ Decays}

We now apply the results of the previous sections to charmless 
$\overline B^0\to\Lambda\bar p\pi^+,\,\Sigma^0\bar p\pi^+,\,p\bar p\overline K^0$, 
and $B^-\to p\bar p\pi^-,\,p\bar p K^-$ decays.
These modes are of interest not just because of possibly large rates,
but also by accessibility in detection.

Let ${\cal T}$ denote the part of the decay amplitude ${\cal M}$
that involves the $B\to{\mathbf B}\overline{\mathbf B}^\prime$
transition matrix element 
$\langle{\mathbf B}\overline{\mathbf B}^\prime|V(A)|B\rangle$, and 
${\cal J}$ denote the part that
involves current-produced 
${\mathbf B}\overline{\mathbf B}^\prime$ matrix element 
$\langle{\mathbf B}\overline{\mathbf B}^\prime|V(A)|0\rangle$. 
We have,
\begin{eqnarray}
{\mathcal M}\left(\Lambda(\Sigma^0)\bar p\pi^+\right)
&=&{\mathcal J}\left(\Lambda(\Sigma^0)\bar p\pi^+\right),
\nonumber\\
  {\mathcal M}\left(p\bar p h\right)
&=&{\mathcal J}\left(p\bar p h\right)
  +{\mathcal T}\left(p\bar p h\right),
\nonumber\\
{\mathcal J}(p\bar p\overline K {}^0)&=&{\mathcal J}\left(p\bar p K^-\right),
\end{eqnarray}
where $h=K^{-,0},\,\pi^{-,0}$.
We note that $\Lambda(\Sigma^0)\bar p\pi^+$ modes only have current-produced
contributions ${\mathcal J}$. 
On the other hand,
the $p\bar p\pi^-$ mode is dominated by 
${\mathcal T}(p\bar p\pi^-)$ contributions, as we will see later.
It can be used to extract baryonic transition form factors which can be
applied to $p\bar p K$, $p\bar p \overline K {}^0$ modes via 
Eq.~(\ref{eq:ppB_FF}).
Furthermore, the $p\bar p \overline K {}^0$ and the $p\bar p K^-$ modes
have identical current-produced matrix elements, as one can easily show
by replacing the spectator quark in the $B^-\to K^-$ transition.

\subsection{$\overline B^0\to\Lambda\bar p \pi^+,\,\Sigma^0 \bar p \pi^+$}

By using Eq.~(\ref{eq:factorization}) and equations of motion
we have,  
\begin{eqnarray}
&&{\cal M}(\Lambda(\Sigma^0)\bar p\pi^+)={\cal J}(\Lambda(\Sigma^0)\bar p\pi^+)
\nonumber\\
&&= \frac{G_F}{\sqrt2}
   \langle \pi^+|\bar u\gamma^\mu (1-\gamma_5)b|\overline B{}^0 \rangle 
   \bigg\{ (V_{ub} V_{us}^*a_1-V_{tb} V_{ts}^* a_4)
   \langle {\rm\bf B}_s\bar p\,|\bar s\gamma_\mu(1-\gamma_5) u|0\rangle
\nonumber\\
&&
+2a_6 V_{tb}V_{ts}^* \frac{(p_{\Lambda(\Sigma^0)}+p_{\bar p})_\mu
(p_{\Lambda(\Sigma^0)}+p_{\bar p})^\nu}{m_b-m_u} 
\bigg\langle\Lambda(\Sigma^0)\bar{p}\,\bigg|
\frac{\bar s\gamma_\nu u}{m_s-m_u} +
\frac{\bar s\gamma_\nu\gamma_5 u}{m_s+m_u}\bigg|0\bigg\rangle \bigg\}.
\label{eq:Lambdappi}
\end{eqnarray}
As stated before, only the current-produced part~(${\cal J}$) contributes,
hence it is similar to the $\overline B {}^0\to K^{(*)-} \pi^+$ mode,
where one only has $\overline B {}^0\to\pi^+$ transition
while, analogous to $\Lambda\bar p$ and $\Sigma^0\bar p$, 
the $K^{(*)-}$ is produced by the current \cite{Ali:1998eb}.

The chiral limit of the vector term is protected by baryon mass differences,
while that of the axial term is protected by Eq. (\ref{eq:0toBBPCAC}).
Since the
contribution from the vector current~($V$) does not interfere
with that from the axial current~($A$), the branching fraction is
a simple sum of the two, i.e. 
${\mathcal B}={\mathcal B}_V+{\mathcal B}_A$.

Taking $\phi_3$ (or $\gamma)=54.8^\circ$ from a recent
analysis~\cite{Ciuchini:2000de}, 
we give in
Table~\ref{tab:Lambdappi} the branching fractions for $\overline
B^0\to\Lambda(\Sigma^0) \bar p \pi^+$. 
The first row is
obtained by extending the asymptotic relations to the whole time-like region
and information from 
the nucleon magnetic form factors $G_M^{p,n}$ for $g_A$, 
while we apply $\tilde f_2=\tilde d_2=\tilde z/2$ in the second row.
Note that we do not need $S_{A,V}$.
Our result for the $\Lambda\bar p\pi^+$ mode in the first line is
consistent with that of Ref.~\cite{Cheng:2001tr}.
We concentrate on the $N_c=3$ case, while $N_c=2,\,\infty$
cases are given to indicate possible non-factorizable effects.
Since $\overline B^0\to\Lambda(\Sigma^0) \bar p \pi^+$ are
penguin-dominated processes, the branching fractions are dominated by 
the $a_6$ term of Eq.~(\ref{eq:Lambdappi}).
One can verify this by comparing different $N_c$ cases in 
Tables~\ref{tab:Lambdappi} and~\ref{tab:ai}.
Since $N_c$ dependence is weak in this term, 
we do not expect large non-factorizable contributions.

\begin{table}[t!]
\caption{\label{tab:Lambdappi}
${\mathcal B}(\overline B^0\to\Lambda \bar p \pi^+[\Sigma^0 \bar p \pi^+])$
(in units of $10^{-6}$) as
decomposed into the vector~(${\mathcal B}_V$) and axial
vector~(${\mathcal B}_A$) current contributions. 
The form factor inputs are explained in the text.
}
\begin{tabular}{lcc|cc|cc}
\hline\hline
     & \multicolumn{2}{c|}{$N_c=2$} & \multicolumn{2}{c|}{$N_c=3$} &
     \multicolumn{2}{c}{$N_c=\infty$}\\
     & ${\mathcal B}_V$ & ${\mathcal B}_A$ & ${\mathcal B}_V$ & ${\mathcal B}_A$ &
     ${\mathcal B}_V$ & ${\mathcal B}_A$ \\
\hline
$G_M^{p,n}$ & $0.13~[0.51]$ & $0.07~[0.16]$ & $0.14~[0.56]$ &
$0.08~[0.17]$ &
$0.18~[0.66]$ & $0.10~[0.21]$ \\
%
$F_A,D_A$ &
--- & 0.31~[0.27] & --- & 0.35~[0.30] & --- & 0.42~[0.36] \\
%
%
\hline\hline
\end{tabular}
\end{table}

%
\begin{figure}[t!]
\epsfig{figure=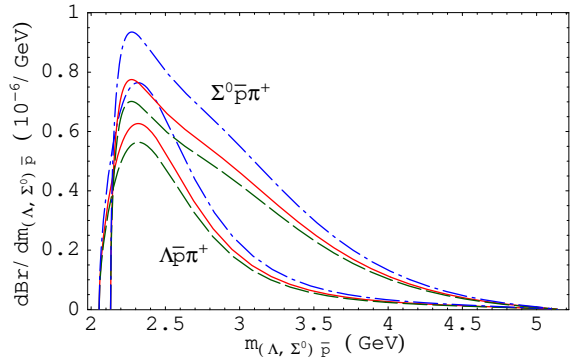,width=3in}
\caption{\label{fig:dGdm_Bsppi} $d{\mathcal B}/dm_{\Sigma^0\bar
p}$~(upper three lines) and $d{\mathcal B}/dm_{\Lambda\bar p}$~(lower
three lines) plots for $\tilde f_2=\tilde d_2=\tilde z/2$. 
Dashed, solid and dot-dashed lines are for $N_c=2,\ 3,\ \infty$, respectively.}
\end{figure}

The axial contribution to $\Lambda (\Sigma^0)\bar p\pi^+$ mode in the second row
is about four (two) times larger than that in the first row.
We find ${\mathcal B}(\Lambda (\Sigma^0)\bar p\pi^+)
\sim [0.5\,(0.9)]\times 10^{-6}$,
giving a larger rate for the $\Sigma^0 \bar p \pi^+$ mode.
These rates are similar to those obtained in \cite{Cheng:2001tr}.

We show in Fig.~\ref{fig:dGdm_Bsppi} the $d{\mathcal B}/dm_{\Lambda\bar p}$ 
and the $d{\mathcal B}/dm_{\Sigma^0\bar p}$ for
the $\tilde f_2=\tilde d_2=\tilde z/2$ case, which gives larger rates. 
One clearly sees threshold enhancement, 
which can be seen as a consequence of 
the need for large-$t$ suppression of the baryon form factors.

\begin{figure}[t!]
\epsfig{figure=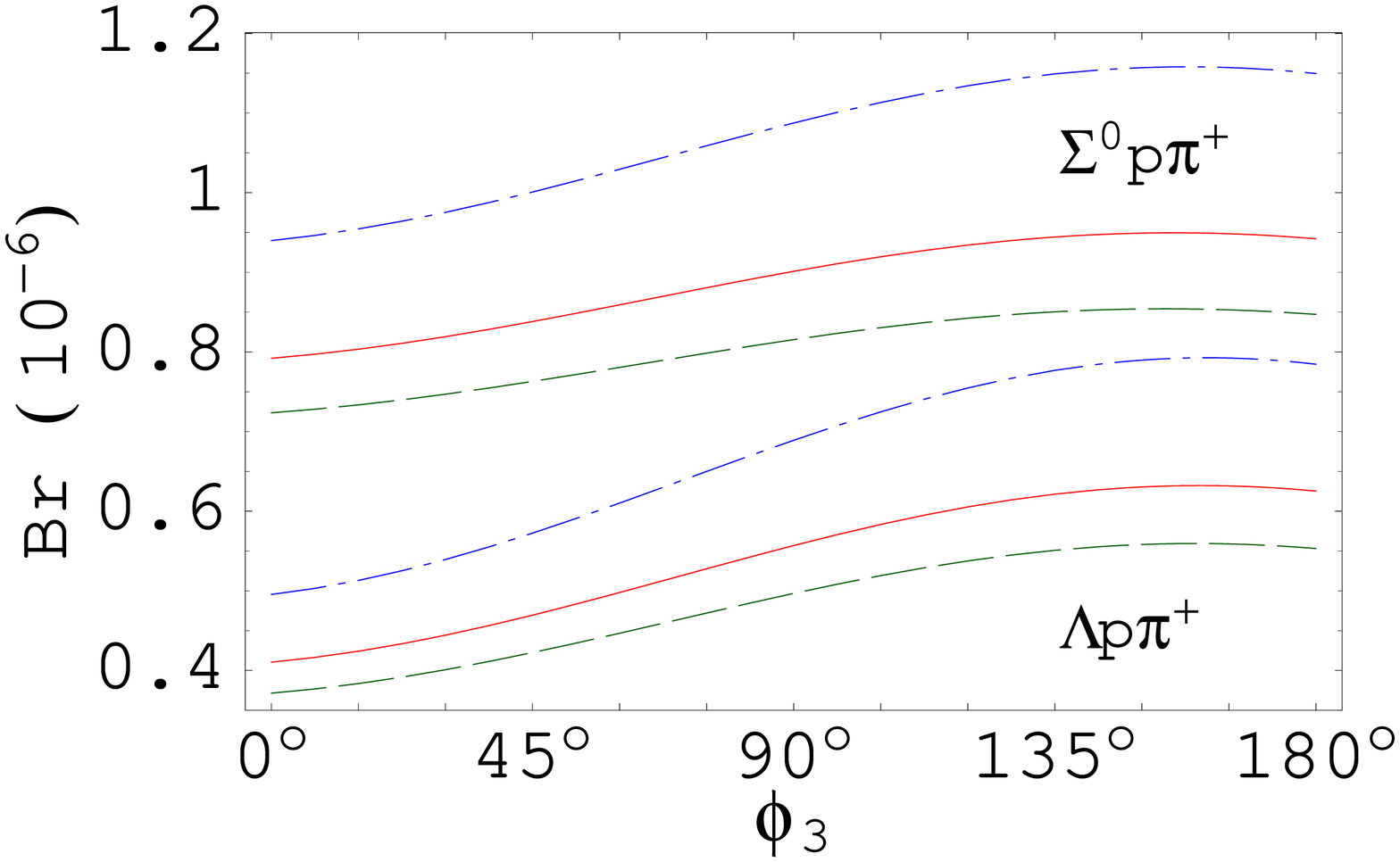,width=3in}
\caption{\label{fig:gamma_Bsppi} $\phi_3$-dependence of
the branching fractions of $\overline B^0\to\Lambda\bar p\pi^+$~(lower three lines) 
and $\Sigma^0\bar p\pi^+$~(upper three lines) for 
$\tilde f_2=\tilde d_2=\tilde z/2$.
Notation the same as Fig.~\ref{fig:dGdm_Bsppi}.}
\end{figure}

Motivated by large $\phi_3$ (or $\gamma$) hints \cite{He:1999mn},
we show the $\phi_3$-dependence of the branching fractions for
$\tilde f_2=\tilde d_2=\tilde z/2$ in Fig.~\ref{fig:gamma_Bsppi}.
The larger rates for larger $\phi_3$ come from tree-penguin interference 
as in the $K^-\pi^+$ case \cite{He:1999mn}.

\subsection{$B\to p\bar p\pi,\,p\bar p K$}

Unlike the $\overline B^0\to\Lambda(\Sigma^0) \bar p\pi^+$ case,
the decay amplitude of $B^-\to p\bar p h^-$ with $h=\pi$ or $K$
contains both the current-produced~(${\cal J}$) and transition~(${\cal
T}$) contributions:
\begin{equation}
{\cal M}(p\bar p h^-)={\cal J}(p\bar p h^-)+{\cal T}(p\bar p h^-)\,,
\end{equation}
where
\begin{eqnarray}
{\cal J}(p\bar p h^-)
&=& \frac{G_F}{\sqrt2}\Biggl\{\langle h^-|(\bar q b)_{V-A}|B^-\rangle
\nonumber\\
&&\times \biggl\langle p\bar p\,\biggl|
V_{ub}V_{uq}^*\,a_2\,\left(\bar uu\right)_{V-A}- V_{tb}V_{tq}^*\biggl[
       a_3\left(\bar uu+\bar dd+\bar ss\right)_{V-A}+
       a_4\left(\bar q q\right)_{V-A}
\nonumber\\
&&+a_5\left(\bar uu+\bar dd+\bar ss\right)_{V+A}
+\frac{3}{2}\,a_9\,\left(e_u \bar uu+e_d \bar dd+e_s \bar
ss\right)_{V-A}\biggr]
\biggr|0\biggr\rangle
\nonumber\\
&&+2 a_6 V_{tb}V_{tq}^*\langle h^-|(\bar q b)_{S-P}|B^-\rangle
              \langle p\bar p\,|\left(\bar q q \right)_{S+P}|0\rangle\Biggr\}
\,, 
\label{eq:J}
\end{eqnarray}
with $q=d$ or $s$ for $h=\pi$ or $K$,
and, from Eq. (\ref{eq:factorization}),
\begin{eqnarray}
{\cal T}(p\bar p h^-)  
&=& \frac{G_F}{\sqrt2}
\bigg\{ 
   (V_{ub} V_{uq}^*a_1-V_{tb} V_{tq}^* a_4)
   \langle h^-|(\bar q u)_{V-A}|0\rangle 
   \langle p\bar p\,|(\bar u b)_{V-A}|B^-\rangle
\nonumber\\
&&-2a_6 V_{tb}V_{tq}^* \frac{m^2_h}{m_b (m_u+m_q)} 
   \langle h^-|(\bar q u)_{V-A}|0\rangle 
   \langle p\bar p\,|(\bar u b)_{V+A}|B^-\rangle 
\bigg\}
\nonumber\\
&=&i\,\frac{G_F}{\sqrt2}f_{h}\,m_b
\biggl[\alpha_{h} \langle p\bar p\,|\bar u b|B^-\rangle
       +\beta_{h} \langle p\bar p\,|\bar u\gamma_5 b|B^-\rangle
\biggr],
\label{eq:T}
\end{eqnarray}
where
\begin{equation}
\alpha_h,\ \beta_h \equiv \left[V_{ub}V_{uq}^*\,a_1-V_{tb}V_{tq}^*
\left(a_4 \pm a_6\,\frac{2\,m_h^2}{m_b(m_q+m_u)}\right)\right]\,.
\label{eq:alphabeta}
\end{equation}

It is of interest to compare the above equations with the familiar 
two-meson decay amplitudes \cite{Ali:1998eb}.
For the $B\to p\bar p$ transition part, the analogous transitions are 
$B^-\to\pi^0,\,\rho^0$ 
(or the isospin-related $\overline B {}^0\to\pi^+,\,\rho^+$).
To single out this effect, 
we can search for two-meson decay modes dominated by such transitions.
For the $\overline B {}^0\to\pi^+,\,\rho^+$ transition dominated modes,
we have $\overline B {}^0\to \pi^+\pi^-(K^-)$ and 
$\overline B {}^0\to \rho^+\pi^-(K^-)$ having decay amplitude
proportional to $\alpha_{\pi(K)}$ and $\beta_{\pi(K)}$, respectively.
For the $B^-\to\pi^0,\,\rho^0$ transitions, we can find $\alpha_{K}$, 
$\beta_{\pi(K)}$ in $B^-\to\pi^0 K^-,\,\rho^0 \pi^-(K^-)$ decay
amplitudes, respectively.
The $\pi^0\pi^-$ mode is different due to the cancellation of strong penguin
in $B^-\to\pi^0$ and $B^- \to\pi^-$ transition parts as they are related 
by isospin.

For the current-produced part, we can find similar terms in 
$B^-\to\pi^0\pi^-$, $\pi^0 K^-$, $\rho^0 \pi^-$ and $\rho^0 K^-$ decay amplitudes.
However, we have additional terms.
The matrix
element of the {\it isosinglet} currents 
$\langle p\bar p|(\bar u u+\bar d d+\bar s s)_{V,A}|0\rangle$ is non-vanishing, 
in contrast to the two-body $K^-\pi^0(\rho^0)$ and $\pi^-\pi^0(\rho^0)$ cases, 
where $\pi^0(\rho^0)$, as a member of an isotriplet, cannot be produced via the
isosinglet current. As we will see,  
$\langle p\bar p|(\bar u u+\bar d d+\bar s s)_{V,A}|0\rangle$ 
give non-negligible contributions to the
$p\bar p K^-(\overline K {}^0)$ modes.

\subsubsection{$B^-\to p \bar p \pi^-$}

%
%

Although we have both current-produced and transition contributions to
the $p\bar p\pi^-$ mode,
the former is expected to be small due to color-suppression
of the $a_2$ tree contribution and the smallness of 
CKM-suppressed penguin contributions, as one can see from Eq.~(\ref{eq:J}). 
We show in Table~\ref{tab:B2pippbar}(a) the contribution from the 
current-produced part ${\mathcal J}(p\bar p\pi^-)$.
As in the previous section, we are interested in the $N_c=3$ case 
and list other $N_c$ cases for estimation of non-factorizable effects.
For $N_c=3$ the main contribution is from the strong penguin terms 
($a_4$, $a_6$), while the tree contribution is small due to
the smallness of $a_2$.
For $N_c=2,\infty$, we have larger $a_2$ and the main contributions
come form the tree amplitude.
With or without non-factorizable parts,
the current-produced contribution is indeed much smaller than the experimental 
rate ${\cal B}\left(p\bar p \pi^-\right) 
= \left(1.9^{+1.0}_{-0.9}\pm0.3\right)\times 10^{-6}$~\cite{Abe:2002ds}.

\begin{table}[t!]
\caption{\label{tab:B2pippbar}
(a) Current-produced ${\mathcal B}_{\mathcal J}(B^-\to p \bar p \pi^-)$ 
in units of $10^{-6}$, and 
(b) strength of transition coefficients $C_{A,P,V5}$ giving rise to
${\mathcal B}_{{\mathcal T}}(B^-\to p \bar p \pi^-)=1.9\times 10^{-6}$. 
}
\begin{center}
\begin{tabular}{lcc|cc|cc}
\multicolumn{7}{c}{(a)~${\cal B}_{{\cal J}}(p \bar p
\pi^-)$ for $\phi_3=54.8^\circ~(90^\circ)$ in units of $10^{-6}$}\\
\hline\hline
     & \multicolumn{2}{c|}{$N_c=2$} & \multicolumn{2}{c|}{$N_c=3$} &
     \multicolumn{2}{c}{$N_c=\infty$}\\

     & ${\mathcal B}_V$ & ${\mathcal B}_A$ & ${\mathcal B}_V$ & ${\mathcal B}_A$
     & ${\mathcal B}_V$ & ${\mathcal B}_A$ \\
\hline
$G_M^{p,n}$ & $0.03~(0.05)$ & $0.03~(0.03)$ & $0.02~(0.03)$ &
$0.01~(0.01)$ & $0.03~(0.03)$ & $0.09~(0.12)$ \\
$F_V,D_V,S_V$ & $0.05~(0.04)$ & --- & $0.00~(0.00)$ & --- & $0.15~(0.10)$ 
& --- \\
$F_A,D_A,S_A$
& --- & $0.09~(0.09)$ & --- & $0.02~(0.03)$ & --- & $0.29~(0.36)$ \\
\hline\hline
\end{tabular}
%
%
\begin{tabular}{lcccccc}
& & & & & & \\
\multicolumn{7}{c}{(b)~$\vert C_X\vert$ values for $\phi_3=54.8^\circ~(90^\circ)$
giving ${\mathcal B}_{{\mathcal T}}=1.9\times 10^{-6}$}\\
\hline\hline
     & \multicolumn{2}{c}{$N_c=2$} & \multicolumn{2}{c}{$N_c=3$} &
     \multicolumn{2}{c}{$N_c=\infty$}\\
     \hline
$|C_A|$ (GeV$^5$)        & \multicolumn{2}{c}{56.61~(63.52)} \
                         & \multicolumn{2}{c}{53.28~(59.86)}
                         & \multicolumn{2}{c}{47.66~(53.66)} \\
$|C_P|$ (GeV$^8$)        & \multicolumn{2}{c}{1233~(1356)} \
                         & \multicolumn{2}{c}{1160~(1278)}
                         & \multicolumn{2}{c}{1038~(1146)} \\
$|C_{V5}|$ (GeV$^5$) \ \ & \multicolumn{2}{c}{57.53~(57.53)} \
                         & \multicolumn{2}{c}{54.11~(54.26)}
                         & \multicolumn{2}{c}{48.36~(48.77)} \\
\hline\hline
\end{tabular}
\end{center}
\end{table}

Since the current-produced part gives small contribution and
the transition part ${\cal T}$ is governed by $a_1$, 
we expect the latter to give major contribution in the $p\bar p\pi$ rate.
The transition part ${\cal T}$ 
involves unknown $B^-\to p\bar p$ transition form factors
$F_{A,P,V5}$ in the matrix elements 
$\langle p\bar p|(\bar u b)_{S,P}|B^-\rangle$.
We illustrate with three cases where only one form factors dominates.
In each case we fit the coefficient $C_A$, $C_P$ or $C_{V5}$ to 
the central value of the experimental measured $p\bar p\pi^-$ rate. 
Note that the matrix elements $\langle p\bar p|(\bar u b)_{S,P}|B^-\rangle$ in
Eq.~(\ref{eq:T}) have nothing to do with the factorized meson $h$, 
hence the obtained coefficients $C_{A,P,V5}$ 
can be applied to the $p\bar p K^-$ mode
(and for the $p\bar p \overline K {}^0$ mode through Eq. (\ref{eq:ppB_FF})) 
as well.

Table~\ref{tab:B2pippbar}(b) shows the obtained values of these coefficients. 
It is interesting to observe that
$C_{A,V5}\sim \left(m_B/2\right)^5$ and $C_P\sim \left(m_B/2\right)^8$. 
Note that the effect of $F_A$ and $F_{V5}$ in the $p\bar p\pi$ decay rate 
are similar.
For $\phi_3=54.8^\circ\,(90^\circ)$, we have
$|\alpha_\pi|\sim (0.9)\,|\beta_\pi|$ for the color allowed tree-dominated part, 
leading to $|C_A|\sim (1.1)\,|C_{V5}|$.
Unlike the current-produced part, 
the transition part is not sensitive to $N_c$,
since $\alpha_\pi$, $\beta_\pi$ (composed of $a_1$, $a_4$ and $a_6$) 
do not depend strongly on $N_c$ as $a_2$. 

\begin{table}[t!]
\caption{\label{tab:Xeq0pppi}
${\cal B}(B^-\to p\bar p \pi^-)$ in
units of $10^{-6}$ for $\phi_3=54.8^\circ~(90^\circ)$ with $\tilde
f_2=\tilde d_2=\tilde z/2$, $s_2=\tilde s_2=0$.}
\begin{ruledtabular}
\begin{tabular}{ccc|cc|cc}
%
& \multicolumn{2}{c|}{$N_c=2$} &
\multicolumn{2}{c|}{$N_c=3$} &
\multicolumn{2}{c}{$N_c=\infty$} \\
   & $+|C_X|$ & $-|C_X|$ & $+|C_X|$ & $-|C_X|$ & $+|C_X|$ & $-|C_X|$ \\
\hline
$X=A$ & $2.24~(2.42)$ & $1.84~(1.66)$ & $1.88~(2.07)$ & $1.96~(1.79)$ &
$1.83~(2.01)$ & $2.85~(2.70)$ \\
$X=P$ & $1.99~(2.15)$ & $2.10~(1.92)$ & $1.86~(2.06)$ & $1.97~(1.80)$ 
& $2.29~(2.50)$ & $2.39~(2.20)$ \\
$X=V5$ & $2.62~(2.55)$ & $1.47~(1.53)$ & $1.99~(2.03)$ & $1.85~(1.82)$ &
$1.41~(1.64)$ & $3.27~(3.06)$ \\
\end{tabular}
\end{ruledtabular}
\end{table}
%

By combining the transition contributions with the current-produced part 
for $\tilde f_2=\tilde d_2=\tilde z/2$, $s_2=\tilde s_2=0$ case, we obtain
the total branching fractions shown in Table~\ref{tab:Xeq0pppi}. 
We see that the rates are still similar to the experimental central value, 
justifying our procedure.

\begin{figure}[t!]
\begin{center}
\epsfig{figure=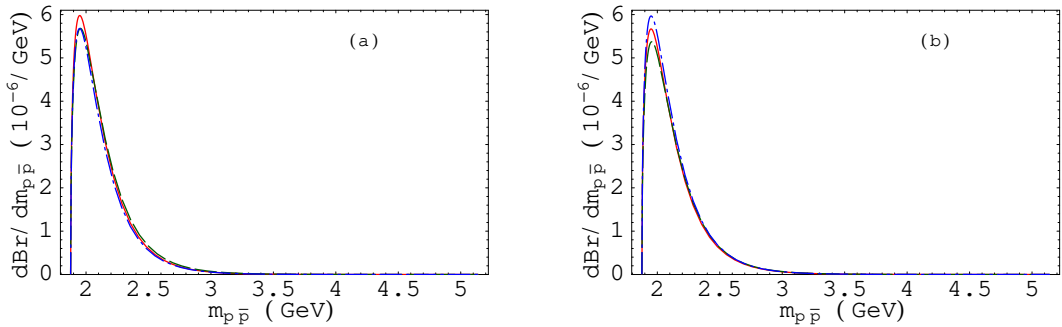,width=6in}
\caption{\label{fig:dGdm_pppi}
$d{\mathcal B}/d\,m_{p\bar p}$ spectrum for the
$B^-\to p\bar p \pi^-$ mode with $N_c=3$ and $\phi_3=54.8^\circ$.
Solid, dash and dot-dash lines in (a) and (b)
correspond to $\mp|C_V|$, $\mp|C_A|$ and $\mp|C_P|$, respectively.}
\end{center}
\end{figure}

In Fig.~\ref{fig:dGdm_pppi} we plot $d{\mathcal B}/d\,m_{p\bar p}$
for the $B^-\to p\bar p \pi^-$ mode, for the $N_c=3$ case 
with $\phi_3=54.8^\circ$,
$C_A=\mp|C_A|$, $C_P=\mp|C_P|$ and $C_{V5}=\mp|C_{V5}|$, respectively.
It is clear that the three cases give close to identical results. 
The threshold enhancement phenomena is evident, as has already been shown
in the $\overline B^0\to \Lambda(\Sigma^0)p\pi^+$ cases.
However, we have a much faster $1/m_{p\bar p}$ suppression here
due to the $1/t^{3,4}$ behavior of the dominant transition form factors,
while for the $\Lambda(\Sigma^0)p\pi^+$ modes the form factors
only behave as $1/t^2$ in the large $t$ limit.
It would be interesting to verify the faster $1/m_{p\bar p}$ 
fall off experimentally.

\begin{figure}[t!]
\epsfig{figure=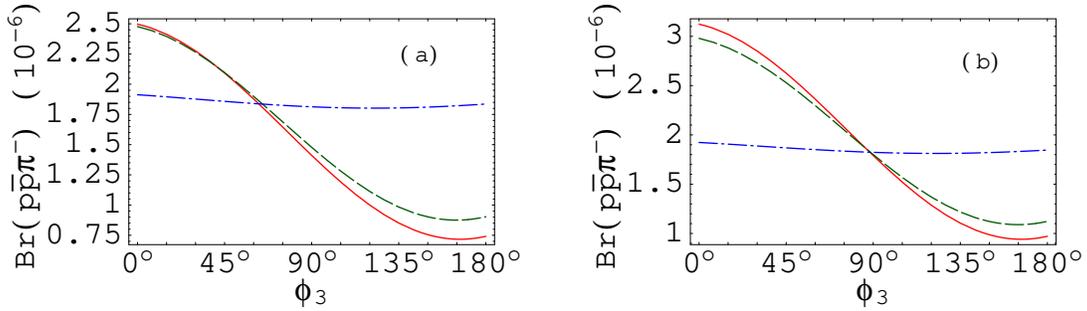,width=6in}
\caption{\label{fig:gamma_pppi} 
$\phi_3$-dependence of $B^-\to p\bar p\pi^-$ branching fraction
for $\tilde f_2=\tilde d_2=\tilde z/2$, $s_2=\tilde s_2=0$ and $N_c=3$.
Solid, dash and dot-dash lines for $-|C_A|$, $-|C_P|$ and $-|C_{V5}|$, 
respectively.
The plots are fixed to $\phi_3=$ (a) $54.8^\circ$,
(b) $90^\circ$ values given in Table~\ref{tab:Xeq0pppi}.}
\end{figure}
%

In Fig.~\ref{fig:gamma_pppi} we illustrate the $\phi_3$-dependence of the 
$p\bar p \pi^-$ branching fractions, with $\phi_3 = 54.8^\circ$ and $90^\circ$ 
results fixed to those of Table~\ref{tab:Xeq0pppi}.
Since this mode is dominated by the transition part,
we expect similar $\phi_3$ behavior as in 
$\overline B {}^0\to \pi^+\pi^-,\,\rho^+\pi^-$ decay rates.
The behavior of the rates from the $C_A$, $C_P$ terms are similar to the 
$\pi^+\pi^-$ case \cite{He:1999mn} rather than the $\pi^-\pi^0$ case.
We have a similar $\alpha_\pi$ term in the $\pi^+\pi^-$ amplitude,
while due to cancellation in the current-produced and the transition terms,  
there is no strong penguin in the $\pi^-\pi^0$ amplitude,
resulting in small tree-penguin interference.
The $C_{V5}$ case is similar to $\overline B {}^0\to \rho^+\pi^-$
and does not show strong $\phi_3$ dependence.
As noted before, both features can be understood from the expression 
of $\alpha_\pi$ and $\beta_\pi$ in Eq. (\ref{eq:alphabeta}).

\subsubsection{$B^-\to p\bar p K^-$}

\begin{table}[t!]
\caption{\label{tab:B2Kppbar}
(a) Current ${\mathcal B}_{{\cal J}}(B^-\to p \bar p K^-)$ and 
(b) transition ${\mathcal B}_{{\cal J}}(B^-\to p \bar p K^-)$
 in units of $10^{-6}$.
} 
\begin{tabular}{lcc|cc|cc}
\multicolumn{6}{c}{(a)~${\cal B}_{{\cal J}}(p \bar p K^-)$ 
in units of $10^{-6}$ for $\phi_3=54.8^\circ~(90^\circ)$}\\
\hline\hline
     & \multicolumn{2}{c|}{$N_c=2$} & \multicolumn{2}{c|}{$N_c=3$} &
     \multicolumn{2}{c}{$N_c=\infty$}\\

     & ${\mathcal B}_V$ & ${\mathcal B}_A$ & ${\mathcal B}_V$ & ${\mathcal B}_A$
     & ${\mathcal B}_V$ & ${\mathcal B}_A$ \\
\hline
$G_M^{p,n}$ & $0.02~(0.02)$ & $0.10~(0.09)$ & $0.01~(0.01)$ &
$0.06~(0.06)$ & $0.33~(0.31)$ & $0.01~(0.02)$ \\
$F_V,D_V,S_V$ & $0.65~(0.69)$ & --- & $0.26~(0.26)$ & --- & $0.02~(0.03)$ 
& --- \\
$F_A,D_A,S_A$
& --- & $0.23~(0.19)$ & --- & $0.12~(0.12)$ & --- & $0.02~(0.05)$ \\
%
%
\hline\hline
\end{tabular}
%
%
\begin{tabular}{lccc}
 & & & \\
\multicolumn{3}{c}{(b)~${\cal B}_{{\cal T}}(p \bar p K^-)$ 
in units of $10^{-6}$ for $\phi_3=54.8^\circ~(90^\circ)$}\\
\hline\hline
               &{$N_c=2$} &{$N_c=3$} &{$N_c=\infty$}\\
\hline
$C_A\qquad$    &{$2.08~(3.42)$} &{$2.12~(3.48)$} &{$2.18~(3.58)$} \\
$C_P\qquad$    &{$1.83~(2.86)$} &{$1.86~(2.91)$} &{$1.90~(2.99)$} \\
$C_{V5}\qquad$ &{$0.26~(0.20)$} &{$0.23~(0.19)$} &{$0.18~(0.18)$} \\
\hline\hline
\end{tabular}
\end{table}
%

For the $B^-\to p\bar p K^-$ decay, we have both a 
transition~(${\mathcal T}$) part and now a more effective 
current-produced~(${\mathcal J}$) part, as shown in Table~\ref{tab:B2Kppbar}(a).
For the vector part,
the largest contributions come from 
$(a_3+a_5)\langle p\bar p|(\bar uu+\bar dd+\bar ss)_V|0\rangle$ 
and $a_4\langle p\bar p|(\bar s s)_{V}|0\rangle$ in Eq.~(\ref{eq:J}). 
In the first line of the table, 
we have $\langle p\bar p|(\bar s s)_{V,A,S,P}|0\rangle=0$ and the contributions
are mainly from the $a_3+a_5$ term.
For $N_c=3$, $a_3+a_5$ is small, resulting in a small ${\mathcal B}_{V}$.
In the second line, the $a_4\langle p\bar p|(\bar s s)_{V}|0\rangle$
term gives ${\mathcal B_V}\sim 0.3\times 10^{-6}$ and interferes differently
with the previous term for different $N_c$ as $a_3+a_5$ changes sign.
On the other hand, the 
$(a_3-a_5)\langle p\bar p|(\bar uu+\bar dd+\bar ss)_A|0\rangle$ 
term dominates in the axial part.
The dependence on $N_c$ of these contributions can be understood
from the behavior of $a_3-a_5$.

We now turn to the transition ($\mathcal T$) part.
For $\phi_3\sim 54.8^\circ\,(90^\circ)$,
we have $|\alpha_K| \sim |\alpha_\pi|$ $(|\alpha_K| \gtrsim |\alpha_\pi|)$ 
and hence ${\mathcal B}_{\mathcal T}(p\bar p K^-) \sim ({\rm or\ } \gtrsim)\
           {\mathcal B}_{\mathcal T}(p\bar p \pi^-)$
from the $C_{A,P}$ contributions.
On the other hand, $a_4$ and $a_6$ in $\beta_K$ are partially canceled and
the tree contribution is CKM suppressed, 
resulting in $|\beta_K|^2\ll|\alpha_K|^2$.
Hence, the $\langle p\bar p|\bar ub|B^-\rangle$ contribution containing $C_{A,P}$
is much larger than the $\langle p\bar p|\bar u\gamma_5b|B^-\rangle$ 
case coming from $C_{V5}$,
as can be seen from Table~\ref{tab:B2Kppbar}(b).
Note further that the $\phi_3=90^\circ$ case from $C_{A,P}$ gives larger rates,
as should be expected from the analogous $K^+\pi^-$ mode \cite{He:1999mn}.

We combine the current-produced contribution with the
transition part and give the total branching fractions in 
Table~\ref{tab:Xeq0ppK}. 
The results prefer the $\phi_3=90^\circ$ case.
Numbers shown in the first two lines of Table~\ref{tab:Xeq0ppK}
are close to the the
experimental rate ${\cal B}(B^\pm\to p\bar p K^\pm)=
(4.3^{+1.1}_{-0.9}\pm0.5)\times10^{-6}$~\cite{Abe:2002ds}.
We see that, for the $\phi_3=90^\circ$, $N_c=3$ and
$C_A=-|C_A|$ case we have ${\mathcal B}=4.26\times 10^{-6}$,
which is closest to the central value of the
experimental rate.
The value only changes by 10\% as we modify $N_c$ to $2$ or $\infty$.

%
\begin{table}[t!]
\caption{\label{tab:Xeq0ppK}
${\cal B}(B^-\to p\bar p K^-)$ in
units of $10^{-6}$ for $\phi_3=54.8^\circ~(90^\circ)$, 
with $\tilde f_2=\tilde d_2=\tilde z/2$, $s_2=\tilde s_2=0$.
}
\begin{center}
\begin{tabular}{ccc|cc|cc}
%
\hline\hline
 & \multicolumn{2}{c|}{$N_c=2$} &
\multicolumn{2}{c|}{$N_c=3$} &
\multicolumn{2}{c}{$N_c=\infty$} \\
   & $+|C_X|$ & $-|C_X|$ & $+|C_X|$ & $-|C_X|$ & $+|C_X|$ & $-|C_X|$ \\
\hline
$X=A$ & $2.56~(3.85)$ & $3.36~(4.75)$ & $2.18~(3.45)$ & $2.82~(4.26)$ &
$2.06~(3.36)$ & $2.38~(3.96)$ \\
$X=P$ & $2.72~(3.75)$ & $2.70~(3.73)$ & $2.25~(3.30)$ & $2.23~(3.28)$ 
& $1.95~(3.08)$ & $1.94~(3.07)$ \\
$X=V5$ & $0.83~(1.08)$ & $1.45~(1.08)$ & $0.42~(0.57)$ & $0.80~(0.56)$ &
$0.16~(0.16)$ & $0.28~(0.36)$ \\
\hline\hline
\end{tabular}
\end{center}
\end{table}
%

\begin{figure}[t!]
\begin{center}
\epsfig{figure=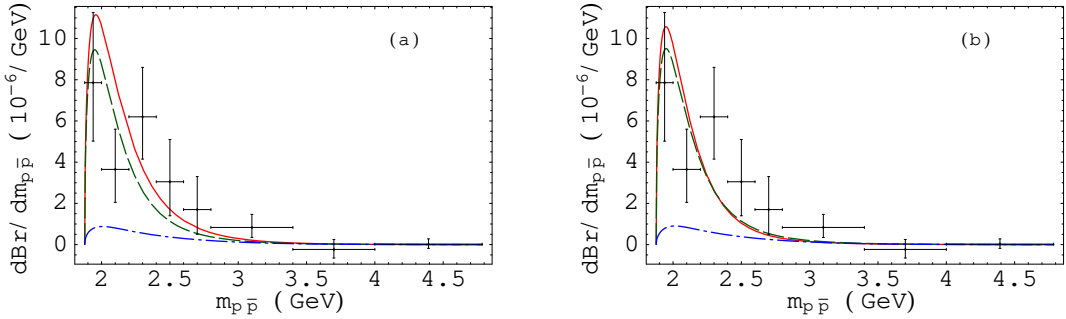,width=6in}
\caption{\label{fig:dGdm_ppKCharged}$d{\mathcal B}/d\,m_{p\bar p}$ 
of the $p\bar p K^-$ mode with $\phi_3=90^\circ$ and $N_c=3$. 
The solid, dashed and the dot-dashed lines in (a) and (b) 
stand for the $C_A=\mp|C_A|$, $C_P=\mp|C_P|$ and the
$C_{V5}=\mp|C_{V5}|$ case, respectively.} 
\end{center}
\end{figure}

We plot $d{\mathcal B}/d\,m_{p\bar p}$ in Fig.~\ref{fig:dGdm_ppKCharged} for 
the $\phi_3=90^\circ$, $N_c=3$ case of Table~\ref{tab:Xeq0ppK}.
The curves from $C_A$, $C_P$ terms are close to data points taken 
from Ref. \cite{Abe:2002ds}. 
The curve from $C_{V5}$ term is too low as expected from the smallness
of $\beta_K$.
For the first two cases, one can see that, 
except for a possible bump at $m_{p\bar p}\sim2.2$--2.4~GeV, the
behavior of the decay spectrum including threshold enhancement
can be explained naturally.
Comparing with Fig.~\ref{fig:dGdm_pppi},
we note that the suppression for large $m_{p\bar p}$ is milder
than the $p\bar p\pi^-$ case.
This is due to the presence of the current-produced part which has
less suppressed form factors ($1/t^2$), which dominates over the
transition part in the large $t$ region.

\begin{figure}[t!]
\epsfig{figure=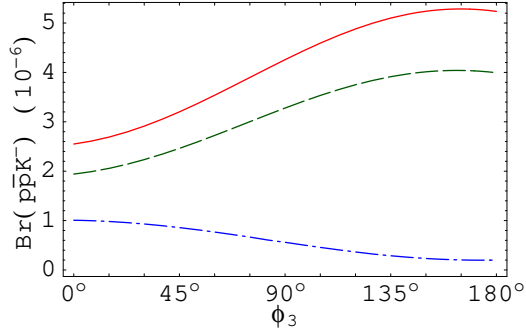,width=3in}
\caption{\label{fig:gamma_ppK} 
$\phi_3$-dependence of ${\cal B}(B^-\to p\bar pK^-)$.
Solid, dash and dot-dash lines stand for the 
$C_A=-|C_A|$, $C_P=-|C_P|$ and $C_{V5}=-|C_{V5}|$ cases, respectively.}
\end{figure}

We give in Fig.~\ref{fig:gamma_ppK} the $\phi_3$-dependence of
${\mathcal B}(p\bar p K^-)$.
The $C_{A,P}$ cases are similar to the $K^-\pi^+$ mode \cite{He:1999mn},
while the $C_{V5}$ case is similar to the $K^-\rho^+$ case
as discussed before. 
We note that the experimental indication that the $p\bar p K^-$ rate is 
larger than the $p\bar p \pi^-$ rate seems to favor 
larger $\phi_3$ values such as $90^\circ$ case,
analogous to $B$ to two meson decay situation \cite{He:1999mn}.

%
%

\subsubsection{$\overline B^0\to p\bar p \overline K {}^0$}

For the $\overline B^0\to p\bar p \overline K {}^0$ decay, we have 
${\cal M}(p\bar p \overline K{}^0)
={\cal J}(p\bar p\overline K {}^0)+{\mathcal T}(p\bar p\overline K {}^0)$.
Since 
$\langle \overline K{}^0|(\bar s b)_{V-A}|\overline B{}^0\rangle 
=\langle K^-|(\bar s b)_{V-A}|B^-\rangle$ by isospin, we have
\begin{equation}
{\cal J}(p\bar p\overline K {}^0)={\cal J}(p\bar p K^-)\,. \label{eq:KeqK0}
\end{equation}
On the other hand,
\begin{equation}
{\cal T}(p\bar p \overline K {}^0)  
= i\,\frac{G_F}{\sqrt2}f_{\overline K {}^0}\,m_b
\biggl[\alpha_{\overline K {}^0} \langle p\bar p\,|\bar d b
                                                  |\overline B {}^0\rangle
       +\beta_{\overline K {}^0}\langle p\bar p\,|\bar d\gamma_5 b
                                                  |\overline B {}^0\rangle
\biggr],
\end{equation}
with
\begin{equation}
\alpha_{\overline K {}^0},\ \beta_{\overline K {}^0}
\equiv -V_{tb}V_{ts}^*
\left(a_4\pm a_6\,\frac{2\,m_{\overline K {}^0}^2}{m_b(m_s+m_d)}\right)\,.
\label{eq:alphabetaK0}
\end{equation}

%
\begin{table}[t!]
\caption{\label{tab:ppbarK0}
(a) Current ${\mathcal B}_{{\cal J}}(\overline B^0\to p \bar p \overline K {}^0)$ 
and
(b) transition ${\mathcal B}_{{\cal T}}(\overline B^0\to p\bar p \overline K{}^0)$
 in units of $10^{-6}$.}
\begin{center}
\begin{tabular}{lcc|cc|cc}
\multicolumn{7}{c}{(a)~${\cal B}_{{\cal J}}(p \bar p\overline K^0)$ 
in units of $10^{-6}$ for $\phi_3=54.8^\circ~(90^\circ)$.}\\
\hline\hline
     & \multicolumn{2}{c|}{$N_c=2$} & \multicolumn{2}{c|}{$N_c=3$} &
     \multicolumn{2}{c}{$N_c=\infty$}\\
     & ${\mathcal B}_V$ & ${\mathcal B}_A$ & ${\mathcal B}_V$ & ${\mathcal B}_A$
     & ${\mathcal B}_V$ & ${\mathcal B}_A$ \\
\hline
$G_M^{p,n}$ & $0.02~(0.02)$ & $0.10~(0.08)$ & $0.01~(0.01)$ &
$0.06~(0.05)$ & $0.31~(0.29)$ & $0.01~(0.02)$ \\
$F_V,D_V,S_V$ & $0.61~(0.65)$ & --- & $0.25~(0.24)$ & --- & $0.02~(0.03)$ 
& --- \\
$F_A,D_A,S_A$
& --- & $0.21~(0.18)$ & --- & $0.12~(0.11)$ & --- & $0.02~(0.05)$ \\
%
%
\hline\hline
\end{tabular}
%
%
\begin{tabular}{lccc}
 & & & \\
\multicolumn{3}{c}{(b)~${\cal B}_{{\cal T}}(p \bar p \overline K {}^0)$ 
in units of $10^{-6}$ for $\phi_3=54.8^\circ~(90^\circ)$.}\\
\hline\hline
     & $N_c=2$ & $N_c=3$ & $N_c=\infty$\\
         \hline
$C_A\qquad$ & $2.74~(3.37)$ & $2.78~(3.43)$ & $2.84~(3.52)$ \\
$C_P\qquad$ & $0.12~(0.14)$ & $0.12~(0.15)$ & $0.13~(0.15)$ \\
$C_{V5}\qquad$ & $1.91~(1.87)$ & $1.93~(1.89)$ & $1.96~(1.95)$ \\
\hline\hline
\end{tabular}
\end{center}
\end{table}
%

We show in Table~\ref{tab:ppbarK0} the separate current-produced and transition
contributions to the $p\bar p \overline K {}^0$ decay rate.
As explained in the above, the current-produced part is identical to the
$p\bar p K^-$ case, except for the difference in $\tau_{B^-}$ and $\tau_{B^0}$.
For the transition part, the transition form factors are related to the
$p\bar p K^-$ case through Eq. (\ref{eq:ppB_FF}).
We concentrate on the $(\bar d b)_S$ form factors instead of $(\bar d b)_P$, 
since $|\alpha_{\overline K {}^0}|\gg |\beta_{\overline K {}^0}|$.
We have ${\cal F}_A=1.1 F_A+0.9 F_{V5}$ and ${\cal F}_P=-F_P/4$
for the $(\bar d b)_S$ form factors.
Since ${\cal F}^{p\bar p\overline K {}^0}_P/{\cal F}^{p\bar pK^-}_P=-1/4$,   
the contribution form the $C_P$ term is very small.
The ratio of the $C_A$ and $C_{V5}$ contributions can be understood as well.
For $\phi_3=54.8^\circ$, the $F_A$ dominated case
is larger than the $F_{V5}$ dominated case by a factor of $11/9$ in amplitude,
giving a rate enhancement $\sim(11/9)^2|C_A/C_{V5}|^2\sim(11/9)^2\sim1.5$.
For $\phi_3=90^\circ$, we have a further 10\% growth in amplitude
due to $|C_A|\sim 1.1 |C_{V5}|$ (as shown in Table~\ref{tab:B2pippbar}(b)),
and the rate enhancement becomes $\sim1.8$.
It is interesting to compare the transition contributions to those in the
$p\bar p K^-$ mode:
for the $C_A$ dominated case, 
${\mathcal B}_{\mathcal T}(p\bar p \overline K {}^0)
\sim {\mathcal B}_{\mathcal T}(p\bar p K^-)$;
for the $C_P$ dominated case, 
${\mathcal B}_{\mathcal T}(p\bar p \overline K {}^0)
\ll {\mathcal B}_{\mathcal T}(p\bar p K^-)$;
for the $C_{V5}$ dominated case, 
${\mathcal B}_{\mathcal T}(p\bar p \overline K {}^0)
\gg {\mathcal B}_{\mathcal T}(p\bar p K^-)$.

We give in Table~\ref{tab:Xeq0ppK0} the full branching fraction by
combining the current-produced and transition parts in amplitude.
For the $N_c=3$, $\phi_3=54.8^\circ\,[90^\circ]$ case, we have
${\mathcal B}(p \bar p\overline K {}^0)=
(0.5$--$3.6)\times 10^{-6}$ [(0.5--$4.3)\times 10^{-6}$].
It could be close to or smaller than the $p\bar p K^-$ rate.

\begin{figure}[t!]
\begin{center}
\epsfig{figure=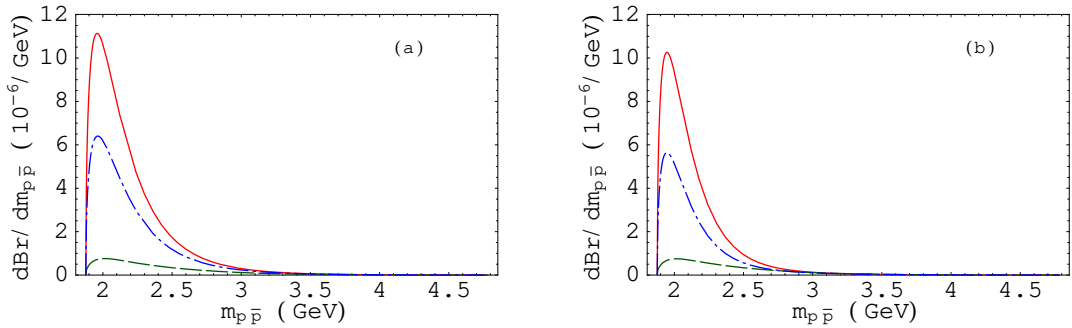,width=6in}
\caption{\label{fig:dGdm_ppK0}
$d{\mathcal B}/d\,m_{p\bar p}$ for the
$\overline B^0\to p\bar p \overline K {}^0$ mode 
with $\tilde f_2=\tilde d_2=\tilde z/2$, $s_2=\tilde s_2$. 
Solid, dash and dot-dash lines in (a) and (b) are for 
$C_A=\mp|C_A|$, $C_P=\mp|C_P|$ and $C_{V5}=\mp|C_{V5}|$ cases, respectively.}
\end{center}
\end{figure}

In Fig.~\ref{fig:dGdm_ppK0} we plot $d{\mathcal B}/d\,m_{p\bar p}$
for the $\overline B^0\to p\bar p \overline K {}^0$ mode.
The decay spectrum for the $C_A$ and $C_{V5}$ cases are similar to 
the $C_A$, $C_P$ cases in the $p\bar p K^-$ mode. 
They exhibit threshold enhancement and a slower fall off for large $t$ 
compared to the $p\bar p\pi^-$ case.
For the $C_A$ case, where the rate is not far from $p\bar p K^-$,
the decay spectrum could be checked soon.

\begin{table}[t!]
\caption{\label{tab:Xeq0ppK0}
${\cal B}(B^-\to p\bar p\overline K {}^0)$ in units of $10^{-6}$,
for $\phi_3=54.8^\circ~(90^\circ)$ 
and $\tilde f_2=\tilde d_2=\tilde z/2$, $s_2=\tilde s_2=0$.
}
\begin{center}
\begin{tabular}{ccc|cc|cc}
%
\hline\hline
 & \multicolumn{2}{c|}{$N_c=2$} &
\multicolumn{2}{c|}{$N_c=3$} &
\multicolumn{2}{c}{$N_c=\infty$} \\
   & $+|C_X|$ & $-|C_X|$ & $+|C_X|$ & $-|C_X|$ & $+|C_X|$ & $-|C_X|$ \\
\hline
$X=A$ & $2.91~(3.53)$ & $4.21~(4.86)$ & $2.68~(3.28)$ & $3.60~(4.28)$ &
$2.71~(3.32)$ & $3.05~(3.88)$ \\
$X=P$ & $0.95~(0.97)$ & $0.94~(0.96)$ & $0.48~(0.50)$ & $0.48~(0.49)$ 
& $0.17~(0.23)$ & $0.16~(0.22)$ \\
$X=V5$ & $2.13~(2.14)$ & $3.33~(3.25)$ & $1.88~(1.85)$ & $2.70~(2.64)$ 
& $1.86~(1.82)$ & $2.14~(2.24)$ \\
\hline\hline
\end{tabular}
\end{center}
\end{table}

\begin{figure}[t!]
\epsfig{figure=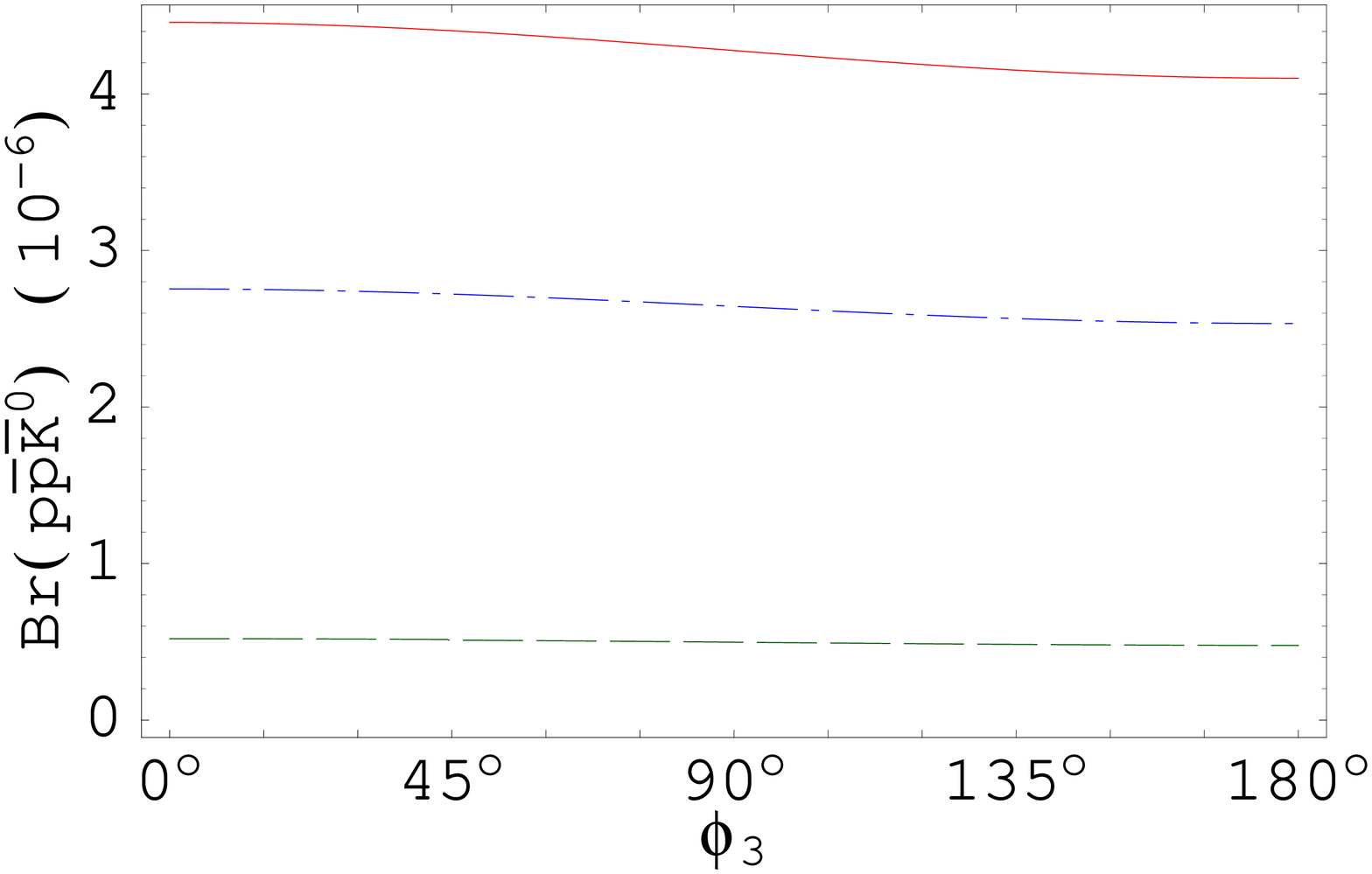,width=3in}
\caption{\label{fig:gamma_ppK0} 
$\phi_3$-dependence of ${\cal B}(\overline B^0\to p\bar p\overline K {}^0)$
for $\tilde f_2=\tilde d_2=\tilde z/2$, $s_2=\tilde s_2=0$. 
Solid, dash and dot-dash lines are for 
$C_A=-|C_A|$, $C_P=-|C_P|$ and $C_{V5}=-|C_{V5}|$ cases, respectively.
}
\end{figure}

%
In Fig.~\ref{fig:gamma_ppK0} we show the
$\phi_3$-dependence of ${\mathcal B}(p\bar p\overline K {}^0)$.
As shown in Eq. (\ref{eq:alphabetaK0}),
the transition part does not have tree-penguin interference and
hence the mild $\phi_3$ dependence is from the sub-dominant 
current-produced term.

%
%

\subsection{Comparison with Other Works}

Before we end this section, we compare our work with some others.
There are approaches that use pole models to evaluate decay matrix elements.
For example, Ref. \cite{Piccinini} use $K^*$ pole for $\bar\Lambda p$ production
in $B\to\eta^\prime\Lambda p$ decay, while Ref. \cite{Cheng:2001tr} use pole models
in two-body and three-body baryonic B decay. We focus on the comparison
with the Ref. \cite{Cheng:2001tr} as we have some subjects in common.

In Ref. \cite{Cheng:2001tr}, Cheng and Yang use a factorization approach in 
the current-produced ($\mathcal J$) amplitude. 
Their approach is similar to ours (up to some technical differences),
hence they obtain results similar to ours in current production dominated modes, 
such as $\bar B^0\to \Lambda \bar p \pi^+$, $\Sigma^0 \bar p\pi^+$.   
However, there is considerable difference in the transition ($\mathcal T$) part.
We factorize the amplitude into a current-produced meson and a
$B$ to baryonic pair transition amplitude. 
In their approach, they use a simple pole model to evaluate this part.
For example, in $B^-\to p\bar p K^-$ decay, they have a strong process
$B^-\to \{\Lambda^{(*)}_b,\Sigma^{0(*)}_b\}\bar p$, 
followed by a weak $\{\Lambda^{(*)}_b,\Sigma^{0(*)}_b\}\to p K^-$ decay.
From their modeling of the strong coupling 
$g_{\Lambda_b\to B^-p}=3\sqrt{3} g_{\Sigma^0_b\to B^-p}$,
they can give $p\bar p K^-$ rate that is close to experimental result 
by using monopole $q^2$ dependence of $g_{\Lambda_b\to B^-p}$.
The $m_{p\bar p}$ spectrum given in Ref.~\cite{Cheng:2001tr} shows a peak around
$t\sim 6\,{\rm GeV}^2$ (or $m_{p\bar p}\sim 2.5$ GeV), 
while we have a sharper peak in lower $m_{p\bar p}$ (around 2 GeV).
The difference is due to the $1/t^3$ behavior in our transition part from
QCD counting rule, while they have $\sim 1/t^2$ from the pole model.
On the other hand, one expects peaking behavior towards large $m_{pK^-}$ 
due to $\Lambda_b$ pole in their approach, 
while in this work we do not expect any structure 
(since $K^-$ and $p$ are factorized)
in the $m_{pK^-}$ spectrum.
In turn, they expect $p\bar p \overline K {}^0$ rate $\sim 10^{-7}$ due to 
the absence of $\Lambda_b$ pole, while we expect a rate that could be
as large as $p\bar p K^-$, although $10^{-7}$ is also possible.
It is up to experiment to check the $m_{pK^-}$ spectrum and 
the $p\bar p \overline K {}^0$ rate. 

We recall that in Ref. \cite{Chua:2001vh} 
we also tried a VMD (dispersion analysis) approach 
\cite{Hammer:1996kx,Mergell:1996bf} in 
the current production dominated $B^0\to D^{*-}p\bar n$ decay. 
In this approach, the strong coupling for each pole is fixed at the pole mass, 
hence each pole gives a monopole contribution to the total form factors. 
One needs to have more than one pole with cancellations in order to 
reproduce the correct QCD counting rule, which is 
$1/t^2$ for current-produced form factors \cite{Hammer:1996kx,Mergell:1996bf}. 
We likely would have the same situation here, that more than one pole is needed to
reproduced the large $t$ behavior.
If we take a multi-pole approach, 
the baryonic transition form factor can be expressed as 
$B\to M_i$ transitions with $M_i$ as one of the mesons, followed by 
a strong process $M_i\to{\bf B\overline B^\prime}$ 
(similar to Ref. \cite{Piccinini}). Summing over $i$, 
the QCD counting rule should be taken as a constraint.  
Instead of doing so, in part because of lack of independent data,
we use a simplified transition form factor motivated from 
the QCD counting rule directly in this work, and wait for experimental data, 
such as  semi-leptonic $B\to{\bf B \bar B^\prime}l\nu$ 
and semi-inclusive $B\to{\bf B \bar B^\prime}X$ 
(similar to $B\to\pi X$ \cite{He:2002ae}), to improve our understanding.
One may resort to the multi-pole approach once these measurements 
become available.

%
\section{Discussion and Conclusion}

In this work we use factorization approach to study charmless
three-body baryonic $B$ decays. We apply SU(3) relations and
QCD counting rules on baryon form factors. 
We identify two mechanisms of baryon pair production, namely
current-produced and transition.
%
The $\Sigma^0\bar p\pi^+$ and $\Lambda\bar p\pi^+$ modes
arise solely from the current-produced part, 
with rates of order $10^{-7}-10^{-6}$.
The $p\bar p\pi^-$, $p\bar p K^-$ and $p\bar p\overline K {}^0$
modes are dominated by transition contributions,
while the current-produced contributions in the last two cases 
are significant.

Due to the absence of $S_{V,A}$ in the current-produced amplitude, 
and the complete absence of the transition amplitude,
$\Sigma^0\bar p\pi^+$ and $\Lambda\bar p\pi^+$ modes are the simplest in this work. 
However, they are sensitive to how we treat the chiral limit of the 
pseudoscalar term (which has an $a_6$ coefficient).
On the other hand, we neglect $F_2$ contribution in the vector part.
It remains to be checked whether this is a good approximation or not. 
It may in turn give us information on $F_2$, or equivalently $G_E$,
from these measurements. In particular, the vector current form factors in
$\Lambda \bar p\pi^+$ are only related to the proton $G_{M,E}$ 
from SU(3) symmetry (as one can see from Eq. (\ref{eq:Lambdappi}) and 
Table~\ref{tab:formfactor}),
and we may obtain information of $G_E^p$ from this mode. 

The $\phi_3$ dependence of $\Lambda \bar p\pi^+$, $\Sigma^0\bar p\pi^+$ rates
are similar to $K^-\pi^+$, $K^{*-}\pi^+$ modes.
Since transition contributions dominate in the $p\bar p h$ modes,
the $\phi_3$ dependence of $p\bar p\pi^-(K^-)$ rate is similar to that in the
two-body $\overline B {}^0\to\pi^+\pi^-(K^-)$ or $\rho^+\pi^-(K^-)$ decays,
while the $\phi_3$ dependence of $p\bar p \overline K {}^0$ rate is mild.

Under factorization, the $p\bar p \pi^-$ and the $p \bar p K^-$ modes have
the same baryonic transition form factors.
Since the $p\bar p \pi^-$ mode is dominated by baryonic transition contribution,
we use it to fit for the transition form factor parameters and 
apply them to the $p \bar p K^-$ case.
To keep $p\bar p\pi^-$ rate around the experimental central value
but allowing $p\bar p K^-$ rate to be larger, 
data seems to favor a larger $\phi_3$.
The $p\bar p \overline K {}^0$ rate can be similar to or much smaller than the
$p\bar p K^-$ rate. 
We do not consider modes involving vector mesons, such as $p\bar p K^*$, 
since they will involve further unknown form factors.

It is interesting that we can reproduce $p\bar pK^-$ decay spectrum based on
QCD counting rules, indicating that the latter is a rather robust theoretical tool.
From QCD counting rules, the current-produced baryonic form factors
behave like $1/t^2$, while the transition baryonic form factors 
behave like $1/t^3$ in large $t\ (\equiv m^2_{\rm\bf B B^\prime})$ limit. 
The $p\bar p\pi^-$ decay is dominated by the transition part and hence
shows a faster damping behavior for large $t$.
In contrast, the $\Lambda \bar p\pi^+$, $\Sigma^0\bar p\pi^+$, $p\bar p K^-$ and
$p\bar p \overline K {}^0$ modes contain current-produced part, and show
a slower damping behavior for large $t$.
We expect an even slower damping behavior (form factors $\sim 1/t$)
in three-body mesonic $B$ decay.
These can be checked soon, especially by comparing the $p\bar p K^-$ and 
the $p\bar p\pi^-$ spectra.

We note that there is a possible bump around $m_{p\bar p}=2.3$ GeV 
in the $p\bar p K^-$ decay spectrum.
It is interesting that the position is close to the mass
of a glueball candidate $\xi(2230)$, also known as $f_J(2220)$,
with $m=2231\pm3.5$ MeV and $\Gamma=23^{+8}_{-7}$ MeV \cite{Groom:in}.
Combining ${\mathcal B}(J/\psi\to\gamma \xi)\gtrsim2.5\times 10^{-3}$
\cite{Groom:in}
and ${\mathcal B}(J/\psi\to\gamma \xi)\,{\mathcal B}(\xi\to p\bar p)
=(1.5^{+0.6}_{-0.5}\pm0.5)\times 10^{-5}$ \cite{Bai:wm},
we have ${\mathcal B}(\xi\to p\bar p)\lesssim6\times 10^{-3}$.
If the rate in the whole $2.2<M_{p\bar p}<2.4$ GeV bin is due to this resonance, 
we would have ${\mathcal B}(B^-\to K^-\xi)\,
 {\mathcal B}(\xi\to p\bar p)\sim 1.24\times10^{-6}$.
We thus get ${\mathcal B}(B^-\to K^-\xi)\gtrsim2\times 10^{-4}$,
or a few times the rate of ${\mathcal B}(B^-\to\eta^\prime K^-)
=(6.5\pm1.7)\times 10^{-5}$~\cite{Groom:in}.
Since both $\xi$ and $\eta^\prime$ are glue rich hadrons, 
while $b\to s$ decays provide glue rich environment~\cite{Atwood:1997de}, 
this may be of great interest.
The underlying dynamics could be $g^*\to g\xi$, which is analogous to
$g^*\to g\eta^\prime$ for $B\to\eta^\prime+X_s$ decay \cite{Atwood:1997bn}.
One should also search in three-body mesonic decay modes.
However, as noted, a slower fall off for non-resonance part, together with
interference with possible nearby resonances, may produce physical background.
But $B\to p\bar p K^-$ decay could be a rather clean mode 
to search for $\xi$ \cite{xi}.

\begin{acknowledgments}
We thank S. J. Brodsky, H.-C. Huang, H.-n. Li and M.-Z. Wang for 
discussions. 
This work is supported in part by the National Science Council of
R.O.C. under Grants NSC-90-2112-M-002-022 and
NSC-90-2811-M-002-038,
the MOE CosPA Project, and the BCP Topical Program of NCTS.
\end{acknowledgments}


\appendix

\section{Some useful Formulas}


In general, for a three-body decay 
$B\to h {\rm\bf B}\overline{\rm\bf B}^{\prime}$, the amplitude can always be
written in the following form:
\begin{eqnarray}
{\cal M}\left(B\to h{\rm\bf B}\overline{\rm\bf B}^{\prime}\right)
&=&
\frac{G_F}{\sqrt{2}}\, \biggl\{ 
{\cal A}\,\bar u(p_{{\mathbf B}})\,/\!\!\!p_h
v(p_{\overline{\mathbf B}^{\prime}})+ {\mathcal B}\,\bar
u(p_{{\mathbf B}})\,/\!\!\!p_h\gamma_5 v(p_{\overline{\mathbf
B}^{\prime}})
\nonumber\\
& &\qquad\qquad
+\,{\cal C}\,
\bar u(p_{{\mathbf B}})\gamma_5
v(p_{\overline{\mathbf B}^{\prime}})
+ {\cal D}\,
\bar u(p_{{\mathbf
B}})v(p_{\overline{\mathbf B}^{\prime}})
\biggr\} \,,\label{eq:general}
\end{eqnarray}
whose absolute square is given by
%
\begin{eqnarray}
\Sigma\,\left|{\cal M}\right|^2&=& G_F^2\,\Biggl\{
\left|{\cal A}\right|^2\biggl[\Bigl(m_B^2+m_2^2-m_{12}^2-m_{23}^2\Bigr)\,
\Bigl(m_{23}^2-m_2^2-m_3^2\Bigr)\nonumber\\
& & \qquad\qquad\qquad\qquad\qquad\qquad\qquad\quad
-m_3^2\,\Bigl(m_{12}^2-(m_1-m_2)^2\Bigr)\biggr]\nonumber\\
&+& 2\,{\rm
Re}\left({\cal
A}\,{\cal D}^*\right)\,
\biggl[m_1\Bigl(m_{23}^2-m_2^2-m_3^2\Bigr)
-m_2
\Bigl(m_B^2+m_2^2-m_{12}^2-m_{23}^2\Bigr)\biggr]\Biggr\}\nonumber\\
&+&\left|{\cal
D}\right|^2\,
\biggl[m_{12}^2-(m_1+m_2)^2\biggr]
\ + \ ({\cal A}\to{\cal B},\,{\cal D}\to {\cal C},\,m_2\to-m_2)\,,
\label{eq:squared}
\end{eqnarray}
where the summation is over all spins, and we have
adopted the convention that the relatively positive baryon
is assigned as particle~1, the other baryon is assigned as
particle~2, the meson $h$ is always assigned as particle~3. 
One can see from the above that only Re$\left({\cal
A}{\cal D}^*\right)$ and Re$\left({\cal B}{\cal C}^*\right)$
appear as interference terms upon squaring.
Given these formulas, the task is now reduced to obtain the 
${\cal A}$--${\cal D}$ terms for an amplitude of interest.

It is straightforward to obtain the decay rate $\Gamma$ 
from the integration of
\begin{equation}
d\Gamma=\frac{1}{\left(2\,\pi\right)^3}\,\frac{1}{32\,m_B^3}
\left(\Sigma\,\left|{\cal
M}\right|^2\right)dm_{12}^2\,dm_{23}^2\,.\label{eq:DecayRate}
\end{equation}



\end{document}